\documentclass[prx,superscriptaddress,twocolumn,longbibliography]{revtex4}

\usepackage{amsmath}
\usepackage{graphicx}
\usepackage{multirow}
\usepackage{subfig}
\usepackage{bbm,color,ulem}
\usepackage{dsfont}
\usepackage{float}
\usepackage{braket}
\allowdisplaybreaks

\newcommand{\ua}{\uparrow}
\newcommand{\da}{\downarrow}

\DeclareMathOperator{\rank}{rank}

\newcommand{\numb}{\addtocounter{equation}{1}\tag{\theequation}}

\begin{document}
\title{Certain exact many-body results for Hubbard model ground states testable in small quantum dot arrays}

\author{Donovan Buterakos}
\author{Sankar Das Sarma}
\affiliation{Condensed Matter Theory Center and Joint Quantum Institute, Department of Physics, University of Maryland, College Park, Maryland 20742-4111 USA}
\date{\today}

\begin{abstract}
	We present several interesting phenomena related to flatband ferromagnetism in the Hubbard model. The first is a mathematical theorem stating certain conditions under which a flatband ferromagnetic must necessarily be degenerate with a nonferromagnetic state. This theorem is generally applicable and geometry-independent, but holds only for a small number of holes in an otherwise filled band. The second phenomenon is a peculiar example where the intuition fails that particles prefer to doubly occupy low-energy states before filling higher-energy states. Lastly, we show a pattern of ferromagnetism which appears in small pentagonal and hexagonal plaquettes at filling factors of roughly 3/10 and 1/4. These examples require only a small number of lattice sites, and may be observable in quantum dot arrays currently available as laboratory spin qubit arrays.
\end{abstract}

\maketitle

\section{Introduction}

The Hubbard model \cite{Hubbard1963} was originally formulated with the goal of explaining the existence of ferromagnetism in common transition metals. The tight-binding Hamiltonian simplistically describes screened electron-electron interactions in a narrow-band metal with the use of only two parameters: the tunneling constant between lattice sites $t_{ij}$, and the onsite Coulomb repulsion energy $U$, as seen in Eq. (\ref{eqn:h1}):

\begin{equation}
H=\sum_{i,j}\sum_{s\in\{\ua,\da\}}-t_{ij}c_{is}^\dagger c_{js}+\sum_i \frac{U}{2}n_i(n_i-1)
\label{eqn:h1}
\end{equation}

where $i$, $j$ are summed over the lattice sites. Despite the simplicity of the Hubbard model, no general solution is known, and while some progress has been made, a complete description of ferromagnetism using this model has not been found after 60 years. However, ferromagnetism has been rigorously proven to occur in the Hubbard model in two very specific instances, namely Nagaoka ferromagnetism and flatband ferromagnetism.

One specific situation where the Hubbard model can be proven to exhibit ferromagnetism is when one electron is added to a half-filled band. This result, proven by Nagaoka, requires that $U=\infty$, $t_{ij}\ge 0$, and that the lattice be sufficiently connected \cite{NagaokaPR1966}. Nagaoka ferromagnetism is a somewhat surprising result when considering that for an exactly half-filled band, the Hubbard model reduces to the Heisenberg model, which generally exhibits antiferromagnetism; yet the addition of a single electron is proven to completely change the magnetic phase. This exact finding of Nagaoka suggests that the Hubbard model, in spite of its apparent simplicity and its seeming dependence on a single effective dimensionless interaction parameter $U/t$, is extremely rich and subtle, since just adding one particle to the half-filled state changes the ground state from being an antiferromagnetic insulator to a ferromagnetic metal. The model is considered to be the paradigmatic model of strong correlations in many body interacting systems, and is foundational in theoretical studies of many phenomena such as Mott transition, ferromagnetism and antiferromagnetism, narrow band systems, high-$T_c$ cuprate superconductivity, spin liquids, etc. Recently, Nagaoka ferromagnetism has been experimentally observed in a small plaquette of four quantum dots which simulates the Hubbard model \cite{DehollainNAT2020}. In an earlier work, we studied Nagaoka ferromagnetism in small quantum dot plaquettes theoretically \cite{ButerakosPRB2019}, connecting with this recent experiment \cite{DehollainNAT2020} and establishing the continued validity of Nagaoka-type ferromagnetism in the Hubbard model even when some of the stringent conditions of the original work \cite{NagaokaPR1966} are relaxed. The idea that the Hubbard model can be simulated by quantum dot plaquettes has been proposed decades ago \cite{StaffordPRL1994,KotlyarPRB1998,Kotlyar2PRB1998}, but current technologies \cite{DehollainNAT2020,vanDiepenPRX2021,MillsNC2019} are capable of realizing these experiments now or in the near future. In fact, the laboratory emulation of the Hubbard model and the associated Mott transition reflected in the observation of the collective Coulomb blockade in a recent experiment on a few quantum dots establish the exquisite control and measurement capability achieved in solid state nanostructures inspired by experimental advances in the semiconductor spin qubit quantum computing platforms \cite{HensgensNAT2017}. Such experiments should enable direct laboratory simulations of the Hubbard model in small lattices of a few electrons.  Although our work is entirely theoretical and quite mathematical in nature, we have been inspired by the rapid recent experimental advance in semiconductor spin qubit platforms consisting of quantum dot arrays, which should enable an experimental verification of our exact (and counter-intuitive) results for the Hubbard model.

In specific cases, it has been proven that the Hubbard model exhibits a different type of ferromagnetism (i.e. distinct from Nagaoka ferromagnetism) known as flat-band ferromagnetism \cite{TasakiPTP1998}. The original proposition for flatband ferromagnetism consisted of a system with a highly degenerate ground state, which could in principle can occur for low-energy (electron) states or high-energy (hole) states. However, due to the Perron-Frobenius theorem, when $t_{ij}$ is positive, the lowest-energy state of a connected lattice is always nondegenerate. Thus flat-band ferromagnetism as mathematically proven in Ref. \onlinecite{TasakiPTP1998} requires degenerate highest-energy states. There has since been much work concerning flatband ferromagnetism \cite{MaksymenkoPRL2012, LeykamAPX2018, DerzhkoPRB2014, GulacsiMPLB2014, AritaPRL2002, SuwaPRB2003, GulacsiPRL2010}. Flatband ferromagnetism has been predicted to occur in armchair graphene nanoribbons \cite{LeeJPCS2009} and twisted bilayer graphene \cite{PonsPRB2020}. It has also been shown that flatband ferromagnetism can occur at half-filling and other filling factors if the flat band occurs in the middle of the spectrum \cite{DerzhkoPRB2014, GulacsiMPLB2014, AritaPRL2002}. However, for this present work, we shall focus on the mathematical formalism originally presented by Tasaki, in which the flat band lies at the top of the spectrum. Because our focus is holes filling in an otherwise filled band, it is convenient to use the particle-hole transformation $\tilde{c}_{is}=c^\dagger_{is}$, where $\tilde{c}_{is}$ is the hole annihilation operator. Under this transformation, Eq. (\ref{eqn:h1}) becomes as follows, up to a constant energy shift:

\begin{equation}
H=\sum_{i,j}\sum_{s\in\{\ua,\da\}}t_{ij}\tilde{c}_{is}^\dagger \tilde{c}_{js}+\sum_i \frac{U}{2}\tilde{n}_i(\tilde{n}_i-1)
\label{eqn:h2}
\end{equation}

The weak flatband ferromagnetism theorem is as follows:

Consider the Hubbard model Eq. (\ref{eqn:h2}) such that the single-hole ground state is $k$-fold degenerate. Then for $h$ holes, with $h\le k$, the ferromagnetic ground state energy will be less than or equal to the energy of any nonferromagnetic state.

The proof of this statement \cite{TasakiPTP1998} follows a similar line of reasoning as Hund's rule. The ferromagnetic ground state consists of the holes filling $h$ of the $k$ degenerate states, and the ground energy corresponds to $h$ times the energy of these states. Because of the Pauli exclusion principle, there will be no doubly-occupied lattice sites, and thus there will be no contribution from the onsite interaction term. Nonferromagnetic states can at best match the same energy, but the onsite interaction term could potentially increase the energy of the states, since doubly-occupied lattice sites are no longer forbidden by the Pauli exclusion principle.

This theorem does not by itself guarantee ferromagnetism, since the inequality between the ferromagnetic ground state energy and the nonferromagnetic energies is not a strict inequality. There have been many works showing that the inequality is strict in certain classes of lattice geometries \cite{MielkeJPA1991,MielkeJPA1992,TasakiPRL1992,TasakiPTP1998,MielkeEPJB2012}; however, it is not known in general for which cases or under what prescribed conditions there will be a unique ferromagnetic ground state.

In this paper we present several results. In Sec. II, we give a mathematical theorem which prescribes certain conditions under which there will be ferromagnetic and antiferromagnetic degeneracy. This theorem is only applicable for a small number of holes, but is independent of geometry, relying only on the number of lattice sites, holes, and flatband single-particle states. In Sec. III, we give an example of a 6-dot configuration where the ground state is ferromagnetic despite not having any degenerate flatband single-particle states. This is, therefore, an example in which ferromagnetism in the Hubbard model can exist without having any band induced single-particle degeneracy. Additionally, the lowest-energy antiferromagnetic state does not doubly-occupy the single-particle ground state as one might expect, instead doubly occupying a higher energy state. In Sec. IV, we present a pattern observed in the ground state energies of various small plaquettes. Specifically, for small plaquettes comprised of hexagons (subsections of hexagonal or diamond lattices) the ground state is ferromagnetic at 1/4 filling. Similarly, for plaquettes consisting of pentagons, the ground state is ferromagnetic at around 3/10 filling. We give a brief explanation for why this may occur, and discuss its applicability to the thermodynamic limit. Finally, in Sec. V, we give our conclusions, briefly noting in particular that the examples we present are experimentally accessible with current quantum dot technologies, each requiring only a few quantum dots. A long Appendix (which can be skipped by the reader) provides many additional results for the experts, and is provided only for the sake of completeness, showing that our theory can be generalized up to 15 dots and many different  plaquette geometries, thus guiding any future experiments on studying quantum ferromagnetism using semiconductor quantum dot spin qubit platforms.  We have many additional results also which are not presented in the current work because the main text and the appendix already provide all representative examples of interest.

\section{Flat Band Ferromagnetic and Nonferromagnetic Degeneracy}

Although it is not known in general which lattice geometries and filling factors are strictly ferromagnetic, it is interesting to ask the converse: are there cases where it is known with certainty that the ferromagnetic ground state is degenerate with nonferromagnetic ground states? This was discussed for certain cases in Ref. \onlinecite{MielkeJPA1999}. We discuss additional cases where this is true and present a general combinatorics-based theoretical result which is completely geometry-independent, although to apply it to specific physical situations requires rather strict conditions on the number of states and holes present in the system. Note that much of the literature about ferromagnetism is concerned with large or infinite lattices. Since we present our results in the context of small quantum dot plaquettes, we are concerned with small, finite systems. For these systems, we define a ``ferromagnetic'' state to be a state with maximal spin (spin $n/2$ if there are $n$ electrons), and an ``antiferromagnetic'' state to be a state with spin 0 or 1/2 (depending on whether there are an odd or even number of fermions).

For the Hubbard model given by Eq. (\ref{eqn:h2}), if there are fewer particles in the system than the number of degenerate single-particle flatband states', then there must be a ferromagnetic ground state (possibly among other ground states) \cite{TasakiPTP1998}. We wish to know in what cases is that ground state is unique. One simple case where flatband ferromagnetic ground states can be shown to be degenerate with nonferromagnetic ground states is when the degenerate flatband states are local. In this case, multiple nonoverlapping local states can be filled with any spin configuration without having any doubly occupied dots. Then the interaction term of the Hamiltonian will not contribute to ground state energy, and thus there will be degenerate ferrommagnetic and nonferromagnetic ground states. For example, in the Kagome lattice, single-hole ground states localize, occupying only six sites comprising one hexagon of the lattice. These local states can be filled with holes without overlap if the filling factor $\nu_h\le1/18$, and thus there will be degenerate ferromagnetic and nonferromagnetic ground states for these filling factors  (note that Mielke proved a stronger statement \cite{MielkeJPA1992} for $\nu_h\le1/6$, but this proof relies strictly on the fact that the Kagome lattice is a line graph, and is not extendable to other lattices). Another example is a small plaquette of dots forming a complete graph ($t_{ij}=t$ for any $i\ne j$). Here the single-hole ground states localize to just two dots, and there is ferromagnetic and nonferromagnetic degeneracy for any $\nu_h\le1/4$.

This effect of nonoverlapping local states leading to ferromagnetic and nonferromagnetic degeneracy is an example of the importance of the so called ``connectivity condition'' to the study of ferromagnetism. The connectivity condition is satisfied only if the flatband states cannot be represented in a basis where they do not overlap with each other. This is often a requisite condition for ferromagnetism to occur without nonferromagnetic degeneracy due to the reason we discussed. However, the connectivity condition is not completely synonymous with flat-band ferromagnetism, as it has been shown that flatband ferromagnetism can still exist in cases when the connectivity condition does not hold due to the behavior of neighboring bands \cite{GulacsiMPLB2014}. Likewise, the connectivity condition by itself does not necessarily guarantee a nondegenerate ferromagnetic ground state when the flat band is not completely filled. For example, consider a the Hubbard model with hopping determined by the edges of the (2,3)-Hamming graph. This system has a flat band consisting of four states. These states must be represented using at least 4 sites each, and there is no way to represent any two of these states using disjoint sets of lattice sites, and thus the connectivity condition is fulfilled. This system does indeed have a ferromagnetic ground state when filled with four particles. However, when this system is occupied by two particles, the ferromagnetic ground state is degenerate with the antiferromagnetic ground state. Thus, there is a way that even completely nonlocal states can still lead to degeneracy. Holes can fill the degenerate flat-band states in a number of ways, each leading to doubly occupied terms $\tilde{c}^\dagger_{i\ua}\tilde{c}^\dagger_{i\da}$ in the full multiparticle many-body state. However, because there are many different ways these states can be filled, it is often possible to form linear combinations in which all doubly occupied terms cancel out. If this is possible, then there will be ferromagnetic and nonferromagnetic ground state degeneracy. In particular, we will show that if the number of distinct ways the flatband states can be filled for a given choice of total spin quantum numbers $s$ and $s_z$ (which we will call $N_f$) is greater than the number of possible doubly occupied terms (which we will call $N_d$), then there will always exist a linear combination of flatband states that cancel out all doubly occupied terms. We will first explicitly calculate the quantities $N_f$ and $N_d$ using combinatorics. We will then give a mathematically rigorous theorem and proof which gives a sufficient condition for degenerate ferromagnetic and nonferromagnetic states.

We calculate the quantities $N_f$ and $N_d$ for a system with $n$ lattice sites, $k$ degenerate flatband states, $h$ holes, and a given total spin $s$. Because spin is conserved, the quantities $N_f$ and $N_d$ will be independent of the choice of $s_z$. We first calculate the total number of multiparticle states $M_f(s_z)$ where each spin-up hole occupies one of the $k$ spin-up flatband states, and each spin-down hole occupies one of the $k$ spin-down flatband states. From basic combinatoric techniques, the number of ways this can occur is simply given by:

\begin{equation}
	M_f(s_z)=\binom{k}{u}\binom{k}{d}
\end{equation}

where $u=h/2+s_z$ is the number of up spins, and $d=h/2-s_z$ is the number of down spins. This quantity $M_f(s_z)$ is related to $N_f$; specifically, it is the sum of $N_f(s)$ for all values of $s$ between $s_z$ and $h/2$.

\begin{equation}
M_f(s_z)=\sum_{s=s_z}^{h/2}N_f(s)
\end{equation}

From this relationship, $N_f(s)$ can be obtained:

\begin{align*}
&N_f(s)=M_f(s_z)-M_f(s_z+1)\\
&=\binom{k}{\frac{h}{2}+s}\binom{k}{\frac{h}{2}-s}-\binom{k}{\frac{h}{2}+s+1}\binom{k}{\frac{h}{2}-s-1}
\numb
\label{eqn:nf}
\end{align*}

Let $N_s(s)$ be the total number of multiparticle states with spin $s$ and a given choice of $s_z$ ($N_s$ is again independent of $s_z$ by symmetry). Then $N_s$ can be calculated in much the same way, except that the number of flatband states $k$ is replaced by the total number of dots $n$:

\begin{equation}
N_s=\binom{n}{\frac{h}{2}+s}\binom{n}{\frac{h}{2}-s}-\binom{n}{\frac{h}{2}+s+1}\binom{n}{\frac{h}{2}-s-1}
\end{equation}

Let $M_u(s_z)$ be the total number of states with $u$ up-spins and $d$ down-spins, such that each hole occupies a unique dot. Then $M_u$ is given by:

\begin{equation}
M_u(s_z)=\binom{n}{u}\binom{n-u}{d}
\end{equation}

Let $N_u(s)$ be the number of states with spin $s$ and a given choice of $s_z$ such that each hole occupies a unique dot. Then like above, $N_u(s)=M_u(s_z)-M_u(s_z+1)$. Then $N_d(s)$, which is the number of states with spin $s$ and a given choice of $s_z$ such that at least one dot is doubly occupied, can be found simply by subtracting $N_u(s)$ from the total number of states $N_s(s)$.

\begin{align*}
&N_d(s)=N_s(s)-N_u(s)\\&=\binom{n}{\frac{h}{2}+s}\binom{n}{\frac{h}{2}-s}-\binom{n}{\frac{h}{2}+s+1}\binom{n}{\frac{h}{2}-s-1}\\&-\binom{n}{\frac{h}{2}+s}\binom{n-\frac{h}{2}-s}{\frac{h}{2}-s}+\binom{n}{\frac{h}{2}+s+1}\binom{n-\frac{h}{2}-s-1}{\frac{h}{2}-s-1}
\numb
\label{eqn:nd}
\end{align*}

Then we have the following theorem:

Consider the Hubbard model Eq. (\ref{eqn:h2}) with $n$ lattice sites such that the single-hole ground state is $k$-fold degenerate. Then for $h$ holes and some total spin $s$, with $2s\le h\le k$, if the quantities given in eqs. (\ref{eqn:nf}) and (\ref{eqn:nd}) satisfy $N_f>N_d$, then there will exist a spin $s$ ground state that is degenerate with the ferromagnetic ground state.

The proof of this theorem is as follows: let $H_T=\sum_{i,j}\sum_{s\in\{\ua,\da\}}t_{ij}\tilde{c}_{is}^\dagger \tilde{c}_{js}$ be the kinetic part of the Hamiltonian, and let $H_U=\sum_i \frac{U}{2}\tilde{n}_i(\tilde{n}_i-1)$ be the interaction term of the Hamiltonian. Restrict the Hilbert space that this Hamiltonian acts on to the set of multiparticle states with $h$ holes, some total spin $s$, and a given choice of $s_z$, and call this Hilbert space $V_s$. Then any state where each dot is occupied no more than once will be an eigenstate of $H_U$ with energy 0, and any state with at least one doubly-occupied dot will be an eigenstate of $H_U$ with nonzero energy. By definition, there will be precisely $N_d$ of these nonzero-energy states, and thus the rank of $H_U$ is equal to $N_d$. Now let $V_f$ be the subspace of $V_s$ where all holes occupy one of the $k$ flatband states, and let $T$ be any linear transformation from $V_f\rightarrow V_s$. Then due to the properties of ranks, $\rank(T^{\dagger}H_UT)\le\rank(H_U)=N_d$. However, since $\dim(V_f)=N_f>N_d$, by the rank-nullity theorem, the nullity of $T^{\dagger}H_UT$ is nonzero, and thus there is a state $\psi_0\in V_f$ which is an eigenstate of $T^{\dagger}H_UT$ with zero energy. Because any state in $V_f$ is an eigenstate of $H_T$ with energy equal to the ferromagnetic ground state energy, then $\psi_0$ will be an eigenstate of $H$ with energy equal to the ferromagnetic ground state energy.

While this theorem is completely general and geometry-independent, in practice the condition $N_f>N_d$ is somewhat difficult to satisfy, and thus the situations where it applies are somewhat limited. However, this theorem simplifies dramatically for the special case where there are only two holes in the system, becoming as follows:

For 2 holes in the Hubbard model given by Eq. (\ref{eqn:h2}) with $n$ lattice sites and $k$ degenerate single-particle ground states, if $k(k+1)/2>n$, then there will exist a spin 0 ground state that is degenerate with the spin 1 ground state.

There are several small configurations of sites which satisfy this condition, including a 4-site complete graph (tetrahedron), the 9-site $(2,3)$-Hamming graph, and a 13-site FCC-sublattice centered around a single site, among other configurations.

It is interesting to see how this theorem scales in the limit where the number of lattice sites $n$ is very large. We will assume in this limit that the number of flatband states $k$ scales proportionally with $n$. For $s=0$, we use Sterling's approximation to simplify the expressions in eqs. (\ref{eqn:nf}) and (\ref{eqn:nd}), and find the following asymptotic behaviors of $N_f$ and $N_d$:

\begin{align*}
N_f&\approx \frac{(k+1)^{\frac{h}{2}+1}k^{\frac{h}{2}-1}}{(\frac{h}{2}+1)!(\frac{h}{2})!}\sim k^h\\
N_d&\approx \frac{(n+1)^{\frac{h}{2}+1}n^{\frac{h}{2}-1}-n^{\frac{h}{2}}(n-\frac{h}{2})^\frac{h}{2}}{(\frac{h}{2}+1)!(\frac{h}{2})!}\sim n^{h-1}
\numb
\end{align*}

Then in the large $n$ limit, the condition $N_f>N_d$ is met for $h<\log n/\log (n/k)$, and thus the maximum number of holes for which this theorem applies scales as $\log$ of the system size. Thus while this theorem is useful for studying small systems of quantum dot plaquettes with just a few sites (since $\log n$ is still comparable in magnitude with $n$ in small systems), as the system size increases the number of holes for which it is valid drops off quickly compared to the total system size. In fact, this serves as an important reminder of the fact that small systems that can be numerically or experimentally simulated cannot always be extrapolated to the thermodynamic limit, as there are effects that are quite strong for small systems which become negligible as the system size increases (in fact, Nagaoka ferromagnetism iself is a dramatic example of this effect also, obviously a filling with one electron or one hole away from half-filling makes no physical sense for a large system although it is a meaningful concept for a small system \cite{ButerakosPRB2019}). Other examples of this include the finite-size effect at the edges of a small lattice, but we have given an example which depends only on the number of states and lattice sites, which stems from the difference between $\log n$ and $n$.

\begin{figure}[!tbh]
	\includegraphics[width=.4\columnwidth]{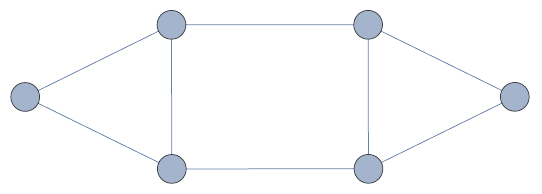}	
	\includegraphics[width=.5\columnwidth]{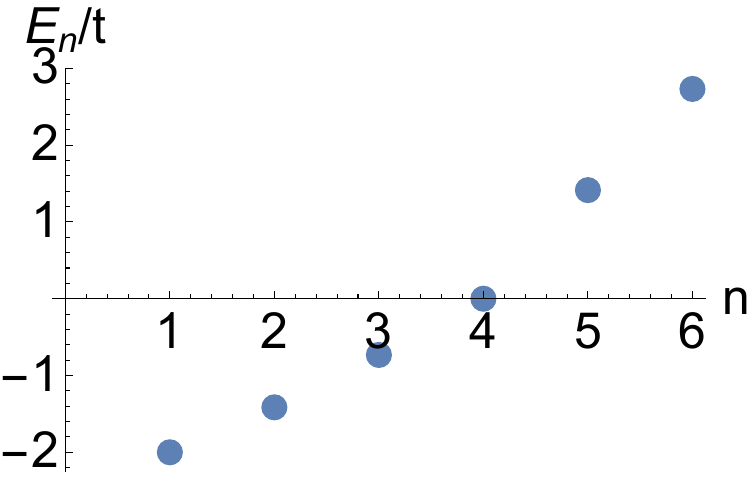}
	\caption{{\bf Left:} 6-dot plaquette which exhibits unusual properties when filled with 3 holes. {\bf Right:} Single-particle energy levels of the plaquette to the left ordered from lowest to highest energy. Note these energies include the effect of the particle-hole transformation, thus corresponding to Eq. (\ref{eqn:h2}).}
	\label{fig:plaquette}
\end{figure}

\section{Example of holes filling higher energy states}

The basic intuition behind Hund's rule is that particles / holes will fill the lowest-energy single-particle states available first, and in the case of degeneracies will form a ferromagnetic spin configuration to avoid energy penalties from double-occupancy of sites. However, in the Hubbard model when $U\gg t$, there are cases where the system breaks this general rule, preferring to fill higher energy single-particle states than necessary in order to eliminate the onsite interaction energy from doubly occupied sites. Thus, strong correlation effects in the Hubbard model explicitly violate Hund's rule leading to counter-intuitive ground states. We give one such example below.

Consider the plaquette of six lattice sites shown in Fig. \ref{fig:plaquette}, with uniform tunneling constant $t$ between nearest neighbors, and $U\gg t$. The single particle-energies for this system are nondegenerate and at a glance are somewhat evenly spaced, as shown in Fig. \ref{fig:plaquette}, although of course there is some anharmonicity. A simple application of Hund's rule would suggest that the ground state should be antiferromagnetic, as holes should attempt to fill the lowest energy state first before beginning to fill the higher energy states, and this is indeed the case when $U$ is small compared to $t$. However when $U$ becomes large, the antiferromagnetic ground state must include some contribution from higher-energy hole configurations, which in some cases can even exceed the energy benefit from doubly occupying the lowest-energy state. This is the case for three holes filling the six-site plaquette discussed above. Exact diagonalization yields a ferromagnetic ground state energy of $E_{3/2}=-4.146t$, whereas the lowest energy antiferromagnetic state has an energy given by:
\begin{equation}
E_{1/2}=-3.949t -1.704t^2/U +O(t^3/U^2)
\end{equation}
 
Thus in the large $U$ limit, the system exhibits a type of ``psuedo-flatband ferromagnetism'' (and not antiferromagnetism) despite the single-particle states being nondegenerate.

\begin{figure}[!tbh]
	\includegraphics[width=.9\columnwidth]{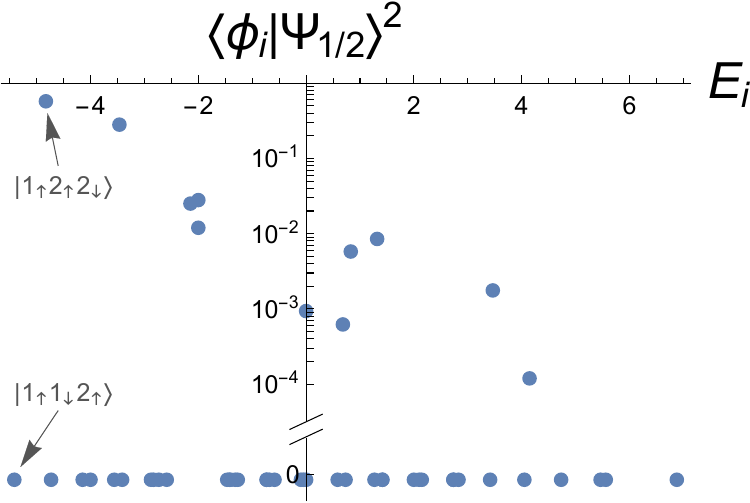}
	\caption{For each product state $\ket{\phi_i}$ of single-particle states, the energy of the state is plotted against the wavefunction overlap of $\ket{\phi_i}$ with the lowest-energy spin 1/2 state $\ket{\Psi_{1/2}}$. States at the bottom of the plot have overlap of exactly 0 because they are protected by symmetry.}
	\label{fig:eoverlap}
\end{figure}

This result becomes even more striking when considering the wavefunction of the lowest-energy antiferromagnetic state $\ket{\Psi_{1/2}}$. Specifically, we consider the product states of single particle eigenstates. These product states, which we denote by $\ket{\phi_i}$ are not necessarily eigenstates themselves, and in general the eigenstates of the system will be linear combinations of the $\ket{\phi_i}$. However, it is often assumed (for example, in Hund's rule) that the ground state is simply the lowest-energy of these product states, and that electrons fill the single-particle states one at a time in order of increasing energy, with perhaps some small correction from interaction effects. However, for this specific configuration, this assumption breaks down entirely. In Fig. \ref{fig:eoverlap}, we plot the energy of each of the product states $\ket{\phi_i}$ versus their overlap with $\Psi_{1/2}$. We find that $\ket{\Psi_{1/2}}$ has an overlap of exactly 0 with the lowest energy product state $\ket{1_\ua1_\da2_\ua}$ (here we have notated the single-hole eigenstates from 1 to 6 in order of increasing energy). Instead, the lowest-energy antiferromagnetic state has a large overlap with the `excited' state $\ket{1_\ua2_\ua2_\da}$ as follows, with the rest of the wavefunction being comprised of higher energy states:
\begin{equation}
\Big|\langle\Psi_{1/2}|1_\ua2_\ua2_\da\rangle\Big|^2=0.573
\end{equation}

The fact that the overlap with $\ket{1_\ua1_\da2_\ua}$ is exactly 0 is due to the $K_4$ symmetry of the Hamiltonian ($H$ is unchanged by $\pi$ rotations about the $x$, $y$, and $z$ axes). The states $\ket{1_\ua1_\da2_\ua}$ and $\ket{1_\ua2_\ua2_\da}$ fall under different symmetry classes, and thus eigenstates of $H$ will not mix these two states. What is surprising is that the system prefers the higher energy of these two states $\ket{1_\ua2_\ua2_\da}$, as this completely contradicts the naive assumption that the system will attempt to fill the lowest-energy single particle states first, only mixing with higher-energy states to eliminate terms corresponding to doubly occupied lattice sites. In this case, the system actually prefers to start with a slightly higher-energy state because it makes the cancellation of doubly-occupied lattice site terms easier. While the symmetry of can account for the fact that many of the $\ket{\phi_i}$ states have zero overlap with $\ket{\Psi_{1/2}}$, we find no simple explanation for why the lowest-energy $\ket{\phi_i}$ state does not have the same symmetry class as $\ket{\Psi_{1/2}}$. For nearly every other cluster this size, the lowest-energy $\ket{\phi_i}$ state has a large overlap with the multi-particle ground state. This example is exceptional because it shows that Hund's rule cannot always be relied upon to gain an understanding of the properties of the multiparticle ground state. This is a subtle and highly nontrivial correlation effect which cannot be captured in terms of any simple general rules based on single particle physics.

\section{Ferromagnetism in pentagonal and hexagonal plaquettes}

We use exact diagonalization methods to find Hubbard model ground states of many different plaquette geometries with up to 16 dots in the infinite $U$ limit. In addition to Nagaoka and flatband ferromagnetism, we also find certain conditions that produce ferromagnetic ground states but do not directly satisfy the Nagaoka or flatband conditions. For example, for a pentagonal plaquettes, we find ferromagnetic ground states when the electron filling factor $\nu_e$ is at or near 3/10, and we find the same for hexagonal plaquettes at 1/4 filling. Here we define a pentagonal (or hexagonal) plaquette as the Hubbard model on a graph with several 5-cycles (6-cycles), but with no cycles of length fewer than 5 (6).

\begin{widetext}
	
\begin{figure}[H]
	\centering
	\includegraphics[width=.3\columnwidth]{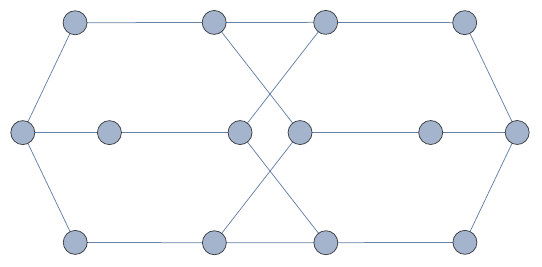}
	\includegraphics[width=.4\columnwidth]{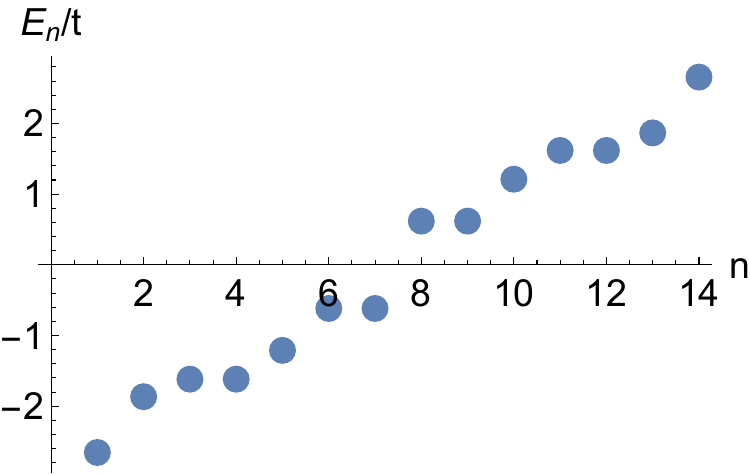}
	\caption{{\bf Left:} Graph of a 14-dot section of a  diamond lattice. {\bf Right:} Single-particle spectrum of the 14-dot diamond lattice plaquette.}
	\label{fig:diamond}
\end{figure}

\begin{table}[H]
	\centering
\begin{tabular}{|c|c|c|c|c|c|c|c|}
	\hline
	\multicolumn{8}{|c|}{14 dots: diamond lattice}\\\hline
	& \multicolumn{7}{|c|}{Spin}\\\hline
	\# of el. & 0, 1/2 & 1, 3/2 & 2, 5/2 & 3, 7/2 & 4, 9/2 & 5, 11/2 & 6, 13/2\\\hline
	13, 15 & -2.593t & -2.599t & -2.6119t & -2.6253t & -2.6339t & -2.6444t & {\bf -2.6554t} \\\hline
	12, 16 & {\bf -4.7895t} & -4.7838t & -4.7671t & -4.7492t & -4.7385t & -4.6294t & -4.5216t \\\hline
	11, 17 & -6.7058t & {\bf -6.7155t} & -6.642t & -6.5838t & -6.567t & -6.1397t &\\\hline
	10, 18 & {\bf -8.3493t} & -8.3128t & -8.2818t & -8.1885t & -8.0084t & -7.7577t &\\\hline
	9, 19 & {\bf -9.2748t} & -9.244t & -9.2344t & -9.0984t & -8.9685t &&\\\hline
	8, 20 & {\bf -9.9614t} & -9.9571t & -9.9493t & -9.9453t & -9.5865t &&\\\hline
	7, 21 & -9.9891t & -10.0439t & -10.1117t & {\bf -10.2045t} &&&\\\hline
	6, 22 & -9.7506t & {\bf -9.7754t} & -9.7284t & -9.5865t &&&\\\hline
	5, 23 & -8.975t & {\bf -9.165t} & -8.9685t &&&&\\\hline
	4, 24 & -7.847t & {\bf -7.9151t} & -7.7577t &&&&\\\hline
	3, 25 & {\bf -6.6649t} & -6.1397t &&&&&\\\hline
	2, 26 & {\bf -4.9985t} & -4.5216t &&&&&\\\hline
\end{tabular}
\caption{Ground state energies of the 14-dot diamond lattice shown in fig. \ref{fig:diamond}, for each number of electrons and spin. The ground state energy for each row is printed in bold. Bold values in the rightmost column represent ferromagnetism.}
\label{tab:diamond}
\end{table}

\end{widetext}

In Table \ref{tab:diamond}, we give the ground state energies of a 14-dot section of a diamond lattice. Energies are given for every value of the total spin and number of electrons. In each row, the ground state is printed in bold, and if this ground state is in the right-most column (corresponding to the maximum value of spin), then the ground state is ferromagnetic. We find that there are two filling factors which produce ferromagnetic ground states (up to particle-hole symmetry). The first is at 13 electrons, which is an example of Nagaoka ferromagnetism. The second is at 7 electrons, corresponding to $\nu_e=1/4$. In Fig. \ref{fig:diamond}, we plot the single particle spectrum in order of increasing energy. Interestingly, the ferromagnetic state at 7 electrons corresponds exactly to a significant gap in the single particle spectrum between the 7th and 8th lowest energy states. In Appendix A, we give many more examples of plaquettes comprised of pentagons or hexagons, and all examples feature ferromagnetic states at a filling near 3/10 or 1/4, and in all cases, this filling corresponds to a significant gap in the single-particle spectrum. We have many such data sets for many different numbers of dots and many different plaquettes, but we restrict ourselves mainly to pentagons and hexagons in the Appendix because the number of finite graphs grows fast with the number of sites, and because pentagons and hexagons most strikingly demonstrate the unexpected flat band ferromagnetism with large gaps which we believe should be of experimental interest in the context of semiconductor spin qubits.

The exact reason why ferromagnetism occurs in these situations is not immediately apparent. It has been proposed that pentagon chains could lead to flat bands which exhibit ferromagnetism at half-filling \cite{AritaPRL2002}. Pentagon chains were specifically chosen such that the Wannier orbitals must overlap, which is necessary to fulfill the local connectivity condition. Satisfying the connectivity condition is nontrivial, and particularly difficult in lattices, which is why pentagon chains were studied \cite{AritaPRL2002}. It was also important that the rings be odd-numbered so that electrons filling the rings would be frustrated. Additionally, the band itself required careful tuning of hopping amplitudes and onsite interaction strengths in order to become flat \cite{AritaPRL2002, SuwaPRB2003, GulacsiPRL2010}. However, the ferromagnetic states we present are almost certainly created by a different mechanism. Our ferromagnetic states are present both in pentagonal and hexagonal plaquettes, the latter of which are bipartite. They also occur at different filling factors (3/10 and 1/4, not at half-filling). Our results are also for small, highly interconnected plaquettes rather than long chains, and we find ferromagnetic states without fine-tuning the Hamiltonian parameters to create a flat band.

We provide a nonrigorous explanation that these states may again be related to the ``psuedo-flatband'' ferromagnetism we discuss above. While the lowest-energy states do not form an exactly flat band, they are perhaps still flat enough to yield ferromagnetic ground states. First, we motivate why a gap appears in the single-particle spectrum around filling factors of 3/10 or 1/4. We note that the single-particle spectrum of the Hubbard model on a 5-dot ring has a fairly large energy gap between the energies $-t\cos 2\pi/5\approx-0.3t$ and $-t\cos 4\pi/5\approx0.8t$. Likewise, the single-particle spectrum of a 6-dot ring has a large gap between the energies $\pm t/2$. Because the plaquettes we study are comprised of several of these connected together, it follows that the states below the gap will hybridize together, as will the states above the gap. Thus we would expect the single-particle spectrum of the composite graph to consist roughly of a group of states with similar energies, followed by a gap, followed by another group of states with similar energies, which is what we observe, though the exact size and location of the gap varies somewhat between examples.

It is possible to see a similar phenomenon to flatband ferromagnetism even if the band is not completely flat. As long as the energy levels in the band are fairly close together, the system will not gain much energy by choosing an antiferromagnetic configuration. However, if there is a large gap between bands, then antiferromagnetic states will have to pay a large energy cost in order to mix with the other band to eliminate doubly-occupied dots. This appears to be the case in the systems which we discussed. However, it is unclear whether the pattern of ferromagnetism at filling factors 3/10 and 1/4 in pentagon or hexagon lattices extends to the thermodynamic limit, as in the thermodynamic limit, energy bands will be continuous, and the explanation we provide will not apply.

\section{Conclusion}

We give examples of several interesting and exact theoretical phenomena in the Hubbard model which require only a few (4-6) lattice sites. The first is a general geometry-independent theorem which has limited applicability (particularly, in the thermodynamic limit) but is mathematically exact. This theorem outlines certain cases of flatband ferromagnetism which necessarily have a nonferromagnetic state that is degenerate with the ferromagnetic ground state, relying only on the number of holes, lattice sites, and degenerate flatband states. The second example is a particular geometric arrangement of six sites which exhibits the unusual behavior where the lowest-energy antiferromagnetic state has no overlap with the product state of the lowest-energy single-particle states, and the many-body ground state is ferromagmmetic rather than antiferromagnetic as implied by Hund's rule. We also show a pattern of ferromagnetic ground states in small pentagonal and hexagonal plaquettes, and discuss its possible relation to imperfect flatband ferromagnetism. Since these phenomena rely only on a small number of sites, these are good candidates to be observed in quantum dot experiments using current technologies. In particular, we believe both can be studied on existing quantum dot arrays associated with spin qubit platforms, directly experimentally establishing nontrivial many-body correlation effects in the Hubbard model ground states.

\acknowledgements

This work is supported by the Laboratory for Physical Sciences.

\appendix

\section{Ground State Energies for Various Quantum Dot Plaquettes}

Below we give the Hubbard model ground state energies for various quantum dot plaquettes of different geometries in the infinite $U$ limit. For every value of total spin, the lowest energy is given in the table, and in each row the ground state is printed in bold.

\begin{widetext}

\subsection{Hexagonal Plaquettes}

\begin{figure}[H]
	\centering
	\includegraphics[width=.3\columnwidth]{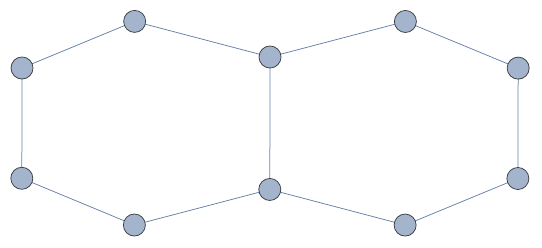}
	\includegraphics[width=.4\columnwidth]{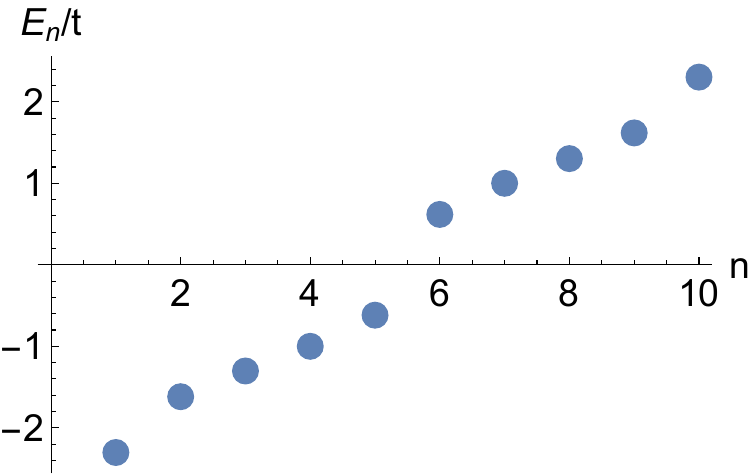}
	\caption{{\bf Left:} Graph of a 10-dot plaquette comprised of 2 adjacent hexagons. {\bf Right:} Single-particle spectrum of the 10-dot hexagon plaquette.}
	\label{fig:fs5}
\end{figure}

\begin{table}[H]
	\centering
\begin{tabular}{|c|c|c|c|c|c|}
	\hline
	\multicolumn{6}{|c|}{10 dots: 2 adjacent hexagons}\\\hline
	& \multicolumn{5}{|c|}{Spin}\\\hline
	\# of el. & 0, 1/2 & 1, 3/2 & 2, 5/2 & 3, 7/2 & 4, 9/2 \\\hline
	9, 11 & -2.293t & -2.293t & -2.298t & -2.297t & {\bf -2.303t } \\\hline
	8, 12 & {\bf -4.093t }  & -4.082t & -4.043t  & -4.016t & -3.921t \\\hline
	7, 13 & -5.579t & {\bf -5.585t }  & -5.487t & -5.224t &\\\hline
	6, 14 & {\bf -6.513t }  & -6.504t & -6.489t & -6.224t &\\\hline
	5, 15 & -6.726t & -6.782t & {\bf -6.842t } &&\\\hline
	4, 16 & -6.366t & {\bf -6.369t }  & -6.224t &&\\\hline
	3, 17 & {\bf -5.517t }  & -5.224t &&&\\\hline
	2, 18 & {\bf -4.16t }  & -3.921t &&&\\\hline
\end{tabular}
\caption{Ground state energies of the 10-dot hexagon plaquette shown in Fig. \ref{fig:fs5}.}
\end{table}

\begin{figure}[H]
	\centering
	\includegraphics[width=.2\columnwidth]{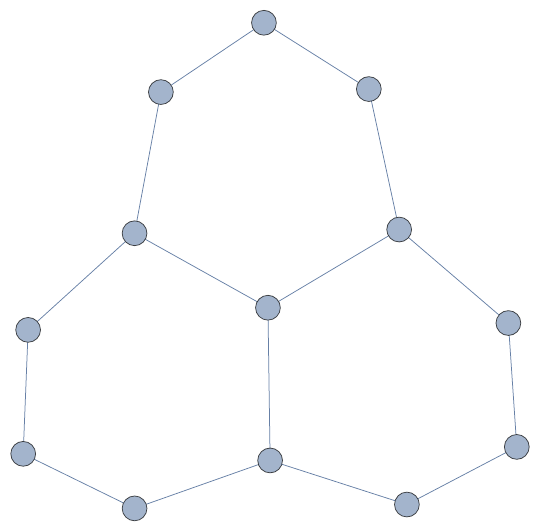}
	\includegraphics[width=.4\columnwidth]{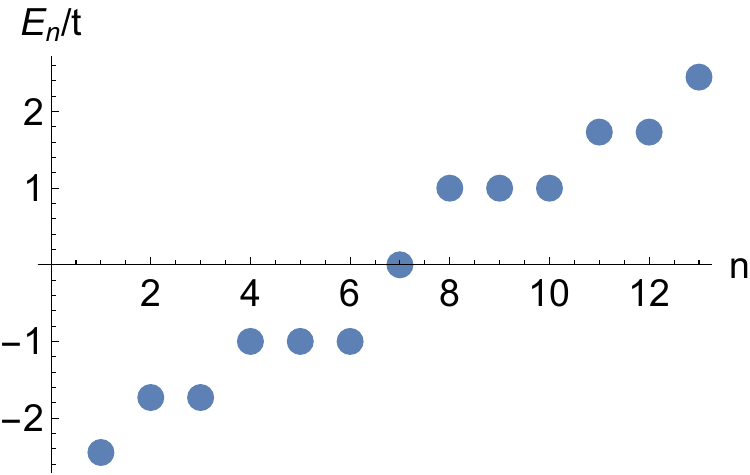}
	\caption{{\bf Left:} Graph of a 13-dot plaquette comprised of 3 adjacent hexagons. {\bf Right:} Single-particle spectrum of the 13-dot hexagon plaquette.}
	\label{fig:f11}
\end{figure}

\begin{table}[H]
	\centering
\begin{tabular}{|c|c|c|c|c|c|c|c|}
	\hline
	\multicolumn{8}{|c|}{13 dots: 3 adjacent hexagons}\\\hline
	& \multicolumn{7}{|c|}{Spin}\\\hline
	\# of el. & 0, 1/2 & 1, 3/2 & 2, 5/2 & 3, 7/2 & 4, 9/2 & 5, 11/2 & 6\\\hline
	12, 14 & -2.4283t & -2.4299t & -2.4355t & -2.4364t & -2.4422t & -2.4434t & {\bf -2.4495t} \\\hline
	11, 15 & {\bf -4.4122t} & -4.4067t & -4.3793t & -4.3515t & -4.33t & -4.1815t &\\\hline
	10, 16 & {\bf -6.1311t} & -6.1271t & -6.1235t & -6.0445t & -6.0155t & -5.9136t &\\\hline
	9, 17 & {\bf -7.515t} & -7.5121t & -7.4809t & -7.256t & -6.9136t &&\\\hline
	8, 18 & {\bf -8.4948t} & -8.4917t & -8.4796t & -8.4616t & -7.9136t &&\\\hline
	7, 19 & -8.8459t & -8.8752t & -8.897t & {\bf -8.9136t} &&&\\\hline
	6, 20 & -8.7807t & -8.7935t & -8.8623t & {\bf -8.9136t} &&&\\\hline
	5, 21 & {\bf -8.341t} & -8.2314t & -7.9136t &&&&\\\hline
	4, 22 & -7.2468t & {\bf -7.4316t} & -6.9136t &&&&\\\hline
	3, 23 & {\bf -6.0194t} & -5.9136t &&&&&\\\hline
	2, 24 & {\bf -4.5183t} & -4.1815t &&&&&\\\hline
\end{tabular}
\caption{Ground state energies of the 13-dot hexagon plaquette shown in Fig. \ref{fig:f11}.}
\end{table}

\begin{figure}[H]
	\centering
	\includegraphics[width=.2\columnwidth]{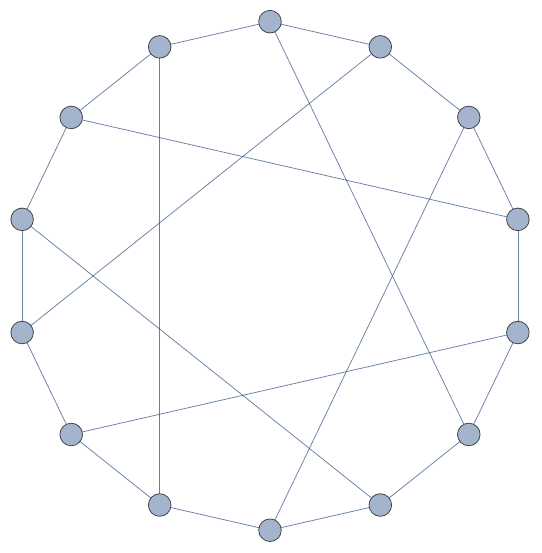}
	\includegraphics[width=.2\columnwidth]{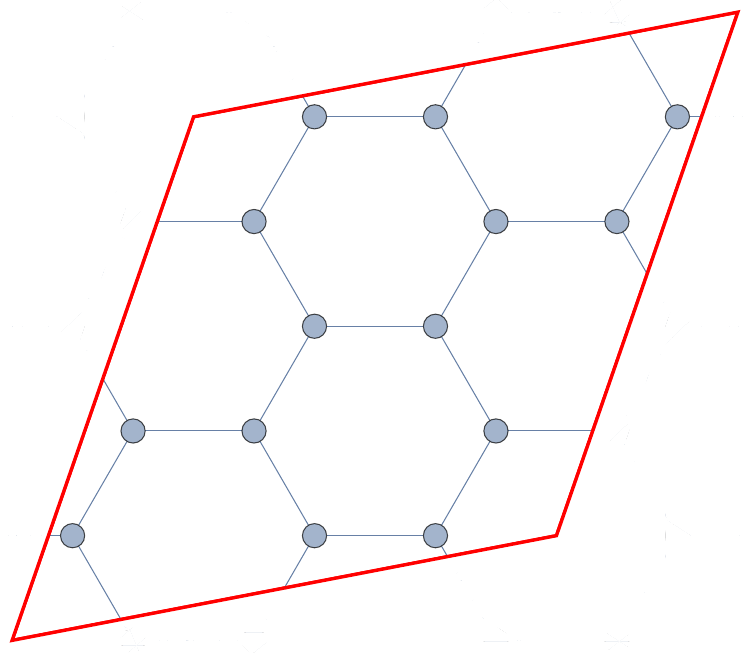}
	\includegraphics[width=.4\columnwidth]{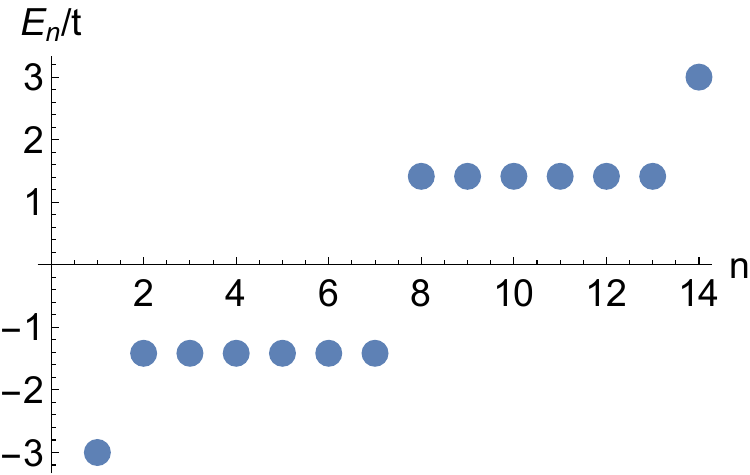}
	\caption{{\bf Left:} Graph of the 14-dot Heawood graph. {\bf Middle:} Heawood graph represented as a hexagon lattice with periodic boundary conditions. {\bf Right:} Single-particle spectrum of the 14-dot Heawood graph plaquette.}
	\label{fig:f16}
\end{figure}

\begin{table}[H]
	\centering
\begin{tabular}{|c|c|c|c|c|c|c|c|}
	\hline
	\multicolumn{8}{|c|}{14 dots: Heawood graph (hexagons with PBC)}\\\hline
	& \multicolumn{7}{|c|}{Spin}\\\hline
	\# of el. & 0, 1/2 & 1, 3/2 & 2, 5/2 & 3, 7/2 & 4, 9/2 & 5, 11/2 & 6, 13/2\\\hline
	13, 15 & -2.8389t & -2.8083t & -2.8259t & -2.8618t & -2.9148t & -2.9477t & {\bf -3t} \\\hline
	12, 16 & {\bf -5.2483t} & -5.2275t & -5.1252t & -5.0448t & -4.9828t & -4.8928t & -4.4142t \\\hline
	11, 17 & {\bf -7.3059t} & -7.2518t & -7.1223t & -6.9363t & -6.5107t & -5.8284t &\\\hline
	10, 18 & {\bf -9.0061t} & -8.9841t & -8.9156t & -8.5495t & -8.0241t & -7.2426t &\\\hline
	9, 19 & -10.0634t & -10.0469t & {\bf -10.0938t} & -9.5609t & -8.6569t &&\\\hline
	8, 20 & -10.8095t & -10.7021t & -10.7533t & {\bf -11.1262t} & -10.0711t &&\\\hline
	7, 21 & -10.7005t & -10.836t & -11.1354t & {\bf -11.4853t} &&&\\\hline
	6, 22 & -9.9128t & -10.0607t & {\bf -10.3206t} & -10.0711t &&&\\\hline
	5, 23 & -9.1832t & {\bf -9.1906t} & -8.6569t &&&&\\\hline
	4, 24 & -8.0058t & {\bf -8.0358t} & -7.2426t &&&&\\\hline
	3, 25 & {\bf -6.8259t} & -5.8284t &&&&&\\\hline
	2, 26 & {\bf -5.6569t} & -4.4142t &&&&&\\\hline
\end{tabular}
\caption{Ground state energies of the 14-dot Heawood graph plaquette shown in Fig. \ref{fig:f16}.}
\end{table}

\newpage

\begin{figure}[H]
	\centering
	\includegraphics[width=.2\columnwidth]{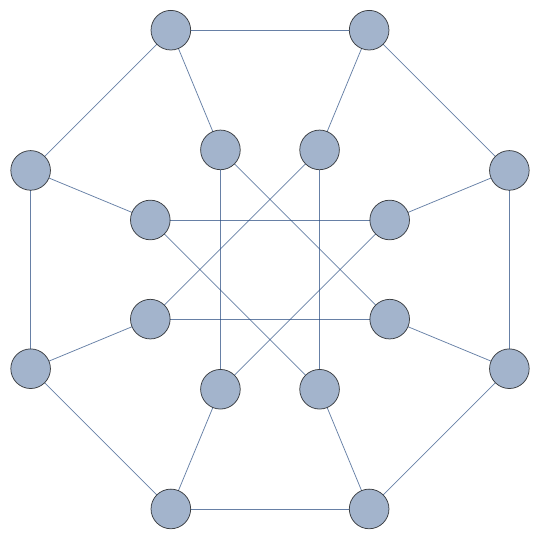}
	\includegraphics[width=.3\columnwidth]{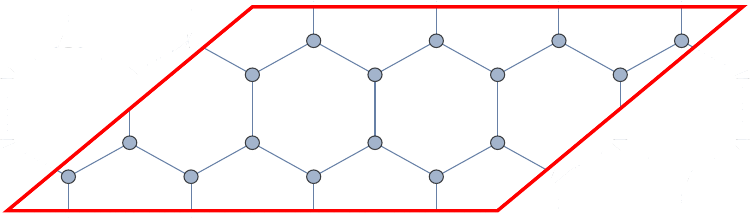}
	\includegraphics[width=.4\columnwidth]{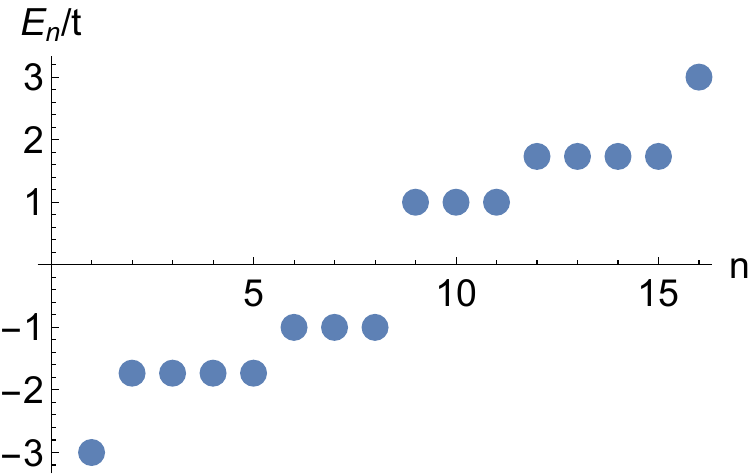}
	\caption{{\bf Left:} Graph of the 16-dot Mobius-Kantor graph. {\bf Middle:} Mobius-Kantor graph represented as a hexagon lattice with periodic boundary conditions. {\bf Right:} Single-particle spectrum of the 16-dot Mobius-Kantor graph plaquette.}
	\label{fig:f23}
\end{figure}

\begin{table}[H]
	\centering
\begin{tabular}{|c|c|c|c|c|c|c|c|c|}
	\hline
	\multicolumn{9}{|c|}{16 dots: Mobius-Kantor graph (hexagons with PBC)}\\\hline
	& \multicolumn{8}{|c|}{Spin}\\\hline
	\# of el. & 0, 1/2 & 1, 3/2 & 2, 5/2 & 3, 7/2 & 4, 9/2 & 5, 11/2 & 6, 13/2 & 7, 15/2\\\hline
	15, 17 & -2.8396t & -2.8379t & -2.8544t & -2.8799t & -2.9059t & -2.9396t & -2.9652t & {\bf -3t} \\\hline
	14, 18 & {\bf -5.3152t} & -5.3044t & -5.2578t & -5.2204t & -5.186t & -5.1218t & -5.0402t & -4.7321t \\\hline
	13, 19 & {\bf -7.4952t} & -7.4692t & -7.406t & -7.3157t & -7.2562t & -6.9284t & -6.4641t &\\\hline
	12, 20 & -9.3443t & {\bf -9.3546t} & -9.3072t & -9.1263t & -8.884t & -8.8116t & -8.1962t &\\\hline
	11, 21 & {\bf -10.8322t} & -10.8302t & -10.8063t & -10.4843t & -10.1118t & -9.9282t &&\\\hline
	10, 22 & {\bf -11.8867t} & -11.7957t & -11.8724t & -11.864t & -11.3916t & -10.9282t &&\\\hline
	9, 23 & -12.2914t & -12.3379t & -12.4179t & {\bf -12.6624t} & -11.9282t &&&\\\hline
	8, 24 & -12.6193t & -12.5361t & -12.5753t & -12.6746t & {\bf -12.9282t} &&&\\\hline
	7, 25 & -11.9385t & -12.0729t & {\bf -12.1114t} & -11.9282t &&&&\\\hline
	6, 26 & -11.1999t & -11.2377t & {\bf -11.5091t} & -10.9282t &&&&\\\hline
	5, 27 & -10.0982t & {\bf -10.1133t} & -9.9282t &&&&&\\\hline
	4, 28 & -8.6982t & {\bf -8.6991t} & -8.1962t &&&&&\\\hline
	3, 29 & {\bf -7.2016t} & -6.4641t &&&&&&\\\hline
	2, 30 & {\bf -5.7139t} & -4.7321t &&&&&&\\\hline
\end{tabular}
\caption{Ground state energies of the 16-dot Mobius-Kantor graph plaquette shown in Fig. \ref{fig:f23}.}
\end{table}

\newpage

\subsection{Pentagonal Plaquettes}

\begin{figure}[H]
	\centering
	\includegraphics[width=.2\columnwidth]{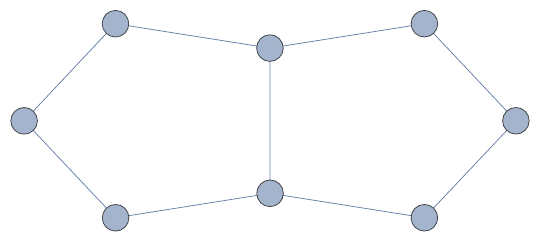}
	\includegraphics[width=.4\columnwidth]{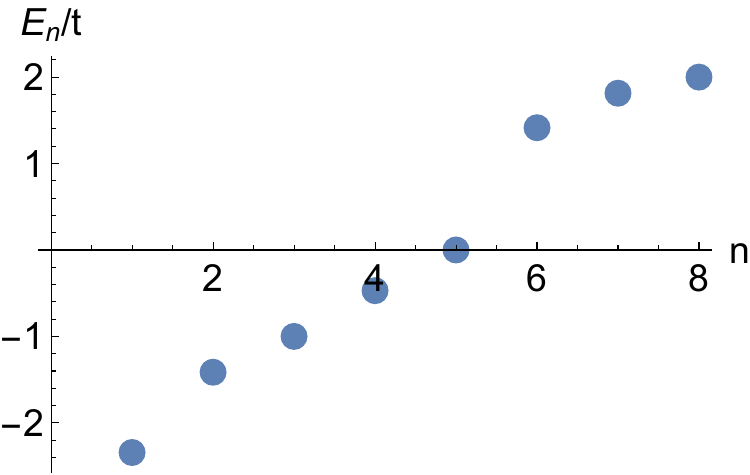}
	\caption{{\bf Left:} Graph of an 8-dot plaquette comprised of 2 adjacent pentagons. {\bf Right:} Single-particle spectrum of the 8-dot pentagon plaquette.}
	\label{fig:fs6}
\end{figure}

\begin{table}[H]
	\centering
\begin{tabular}{|c|c|c|c|c|}
	\hline
	\multicolumn{5}{|c|}{8 dots: 2 adjacent pentagons}\\\hline
	& \multicolumn{4}{|c|}{Spin}\\\hline
	\# of el. & 0, 1/2 & 1, 3/2 & 2, 5/2 & 3, 7/2 \\\hline
	14 & -3.499t & {\bf -3.814t } &&\\\hline
	13 & -5.063t & {\bf -5.228t } &&\\\hline
	12 & {\bf -5.612t }  & -5.547t & -5.228t &\\\hline
	11 & {\bf -5.153t }  & -5.066t  & -4.757t &\\\hline
	10 & {\bf -4.023t }  & -3.945t & -3.878t & -3.757t \\\hline
	9 & -2.299t & -2.329t  & -2.327t & {\bf -2.343t } \\\hline
	8 & 0 & 0  & 0  & 0 \\\hline
	7 & -2.325t & {\bf -2.328t }  & -2.267t & -2t \\\hline
	6 & {\bf -4.033t }  & -4.024t  & -4.006t  & -3.814t \\\hline
	5 & -5.095t & -5.16t & {\bf -5.228t } &\\\hline
	4 & -5.418t & {\bf -5.436t }  & -5.228t &\\\hline
	3 & {\bf -5.102t }  & -4.757t &&\\\hline
	2 & {\bf -4.093t }  & -3.757t &&\\\hline
\end{tabular}	
\caption{Ground state energies of the 8-dot pentagon plaquette shown in Fig. \ref{fig:fs6}.}
\end{table}

\newpage

\begin{figure}[H]
	\centering
	\includegraphics[width=.2\columnwidth]{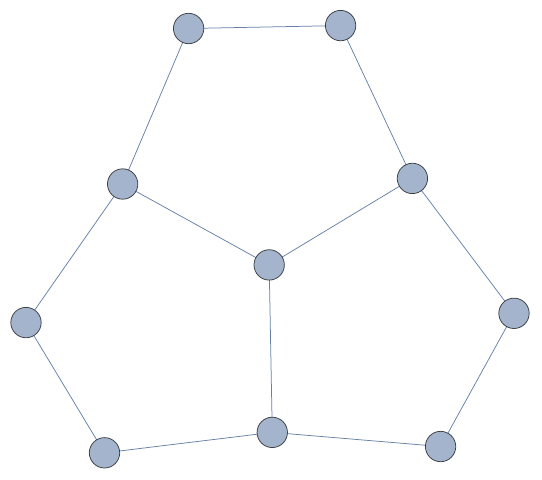}
	\includegraphics[width=.4\columnwidth]{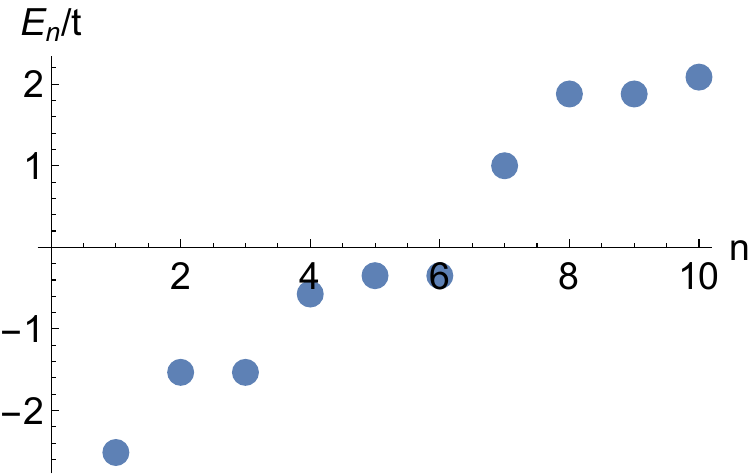}
	\caption{{\bf Left:} Graph of a 10-dot plaquette comprised of 3 adjacent pentagons. {\bf Right:} Single-particle spectrum of the 10-dot pentagon plaquette.}
	\label{fig:fs7}
\end{figure}

\begin{table}[H]
	\centering
	\begin{tabular}{|c|c|c|c|c|c|}
		\hline
		\multicolumn{6}{|c|}{10 dots: 3 adjacent pentagons}\\\hline
		& \multicolumn{5}{|c|}{Spin}\\\hline
		\# of el. & 0, 1/2 & 1, 3/2 & 2, 5/2 & 3, 7/2 & 4, 9/2 \\\hline
		18 & -3.756t  & {\bf -3.966t } &&&\\\hline
		17 & -5.548t  & {\bf -5.845t } &&&\\\hline
		16 & -6.511t  & -6.741t  & {\bf -6.845t } &&\\\hline
		15 & {\bf -7.171t }  & -7.117t  & -6.498t &&\\\hline
		14 & {\bf -6.816t  }  & -6.783t  & -6.457t  & -6.15t &\\\hline
		13 & {\bf -5.888t }  & -5.746t & -5.733t  & -5.578t &\\\hline
		12 & {\bf -4.429t }  & -4.352t   & -4.318t  & -4.268t   & -4.046t \\\hline
		11 & -2.466t  & -2.472t & -2.488t  & -2.496t  & {\bf -2.514t } \\\hline
		10 & 0  & 0  & 0  & 0 &  0 \\\hline
		9 & -2.48t  & {\bf -2.481t }  & -2.451t  & -2.342t  & -2.086t \\\hline
		8 & {\bf -4.424t }  & -4.421t  & -4.41t  & -4.396t  & -3.966t \\\hline
		7 & -5.798t  & -5.827t  & -5.841t  & {\bf -5.845t } &\\\hline
		6 & -6.63t  & -6.657t  & -6.762t  & {\bf -6.845t } &\\\hline
		5 & {\bf -6.984t }  & -6.861t  & -6.498t &&\\\hline
		4 & -6.434t  & {\bf -6.72t  }  & -6.15t &&\\\hline
		3 & {\bf -5.698t }  & -5.578t &&&\\\hline
		2 & {\bf -4.535t }  & -4.046t &&&\\\hline
	\end{tabular}
	\caption{Ground state energies of the 10-dot pentagon plaquette shown in Fig. \ref{fig:fs7}.}
\end{table}

\newpage

\begin{figure}[H]
	\centering
	\includegraphics[width=.2\columnwidth]{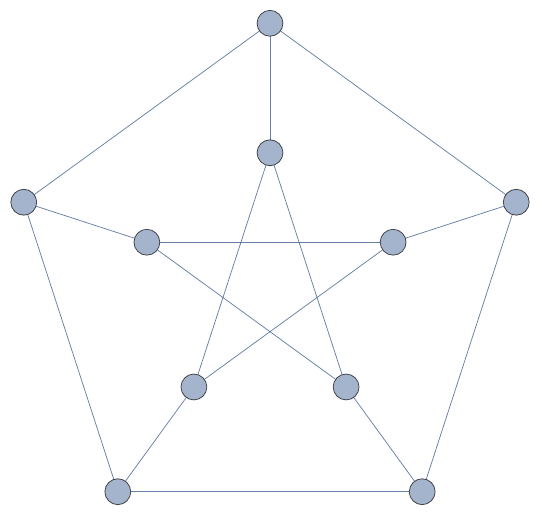}
	\includegraphics[width=.4\columnwidth]{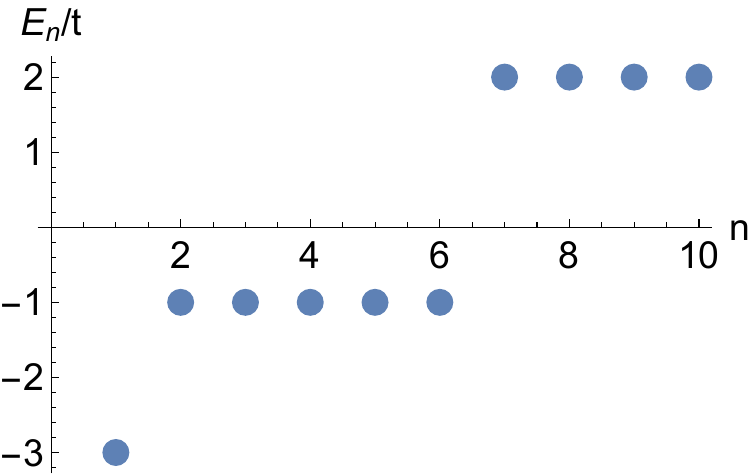}
	\caption{{\bf Left:} Graph of the 10-dot Petersen graph. {\bf Right:} Single-particle spectrum of the 10-dot Petersen graph plaquette.}
	\label{fig:f1}
\end{figure}

\begin{table}[H]
	\centering
\begin{tabular}{|c|c|c|c|c|c|}
	\hline
	\multicolumn{6}{|c|}{10 dots: Petersen graph}\\\hline
	& \multicolumn{5}{|c|}{Spin}\\\hline
	\# of el. & 0, 1/2 & 1, 3/2 & 2, 5/2 & 3, 7/2 & 4, 9/2\\\hline
	18 & -3.9173t & {\bf -4t} &&&\\\hline
	17 & -5.7459t & {\bf -6t} &&&\\\hline
	16 & -6.8988t & -7.2802t & {\bf -8t} &&\\\hline
	15 & -7.8383t & {\bf -7.9503t} & -7t &&\\\hline
	14 & {\bf -7.7187t} & -7.6523t & -6.9835t & -6t &\\\hline
	13 & {\bf -6.708t} & -6.2825t & -5.8116t & -5t &\\\hline
	12 & {\bf -5.1579t} & -4.8585t & -4.8241t & -4.5823t & -4t \\\hline
	11 & -2.6981t & -2.7244t & -2.8192t & -2.8951t & {\bf -3t} \\\hline
	10 & 0 & 0 & 0 & 0 & 0 \\\hline
	9 & {\bf -2.8097t} & -2.7863t & -2.6564t & -2.4142t & -2t \\\hline
	8 & {\bf -4.9536t} & -4.9468t & -4.9152t & -4.5866t & -4t \\\hline
	7 & -6.5726t & -6.5144t & {\bf -6.8102t} & -6t &\\\hline
	6 & -7.3782t & -7.309t & -7.614t & {\bf -8t} &\\\hline
	5 & -7.0921t & {\bf -7.3878t} & -7t &&\\\hline
	4 & -6.7057t & {\bf -6.7932t} & -6t &&\\\hline
	3 & {\bf -6.0931t} & -5t &&&\\\hline
	2 & {\bf -5.4641t} & -4t &&&\\\hline
\end{tabular}
\caption{Ground state energies of the 10-dot Petersen graph plaquette shown in Fig. \ref{fig:f1}.}
\end{table}

\newpage

\begin{figure}[H]
	\centering
	\includegraphics[width=.2\columnwidth]{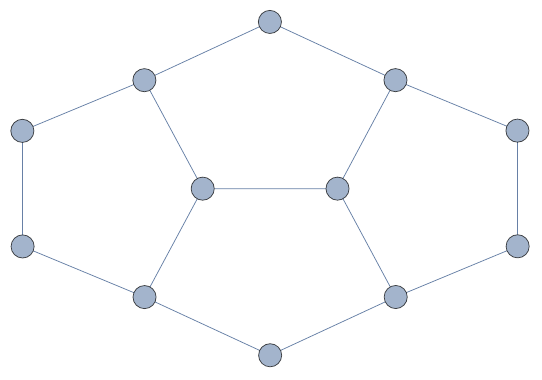}
	\includegraphics[width=.4\columnwidth]{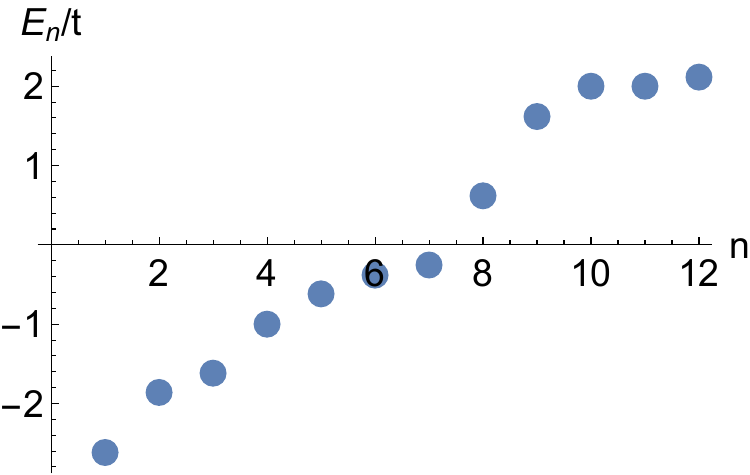}
	\caption{{\bf Left:} Graph of a 12-dot plaquette comprised of 4 adjacent pentagons. {\bf Right:} Single-particle spectrum of the 12-dot pentagon plaquette.}
	\label{fig:f5}
\end{figure}

\begin{table}[H]
	\centering
\begin{tabular}{|c|c|c|c|c|c|c|}
	\hline
	\multicolumn{7}{|c|}{12 dots: 4 adjacent pentagons}\\\hline
	& \multicolumn{6}{|c|}{Spin}\\\hline
	\# of el. & 0, 1/2 & 1, 3/2 & 2, 5/2 & 3, 7/2 & 4, 9/2 & 5, 11/2\\\hline
	22 & -3.9693t & {\bf -4.1149t} &&&&\\\hline
	21 & -5.9239t & {\bf -6.1149t} &&&&\\\hline
	20 & -7.4645t & -7.594t & {\bf -7.7329t} &&&\\\hline
	19 & -8.3466t & {\bf -8.362t} & -8.351t &&&\\\hline
	18 & {\bf -8.7285t} & -8.7119t & -8.6377t & -8.0969t &&\\\hline
	17 & {\bf -8.4649t} & -8.4168t & -8.0788t & -7.7149t &&\\\hline
	16 & {\bf -7.7016t} & -7.6604t & -7.5074t & -7.2633t & -7.0969t &\\\hline
	15 & {\bf -6.4056t} & -6.2942t & -6.2205t & -6.2093t & -6.0969t &\\\hline
	14 & {\bf -4.6774t} & -4.6668t & -4.629t & -4.6165t & -4.6019t & -4.4788t \\\hline
	13 & -2.5629t & -2.5744t & -2.5851t & -2.5981t & -2.6078t & {\bf -2.618t} \\\hline
	12 & 0 & 0 & 0 & 0 & 0 & 0 \\\hline
	11 & {\bf -2.5706t} & -2.5699t & -2.5571t & -2.5202t & -2.3996t & -2.1149t \\\hline
	10 & {\bf -4.7029t} & -4.6957t & -4.6949t & -4.6881t & -4.5666t & -4.1149t \\\hline
	9 & -6.3047t & -6.3257t & -6.3346t & {\bf -6.341t} & -6.1149t &\\\hline
	8 & -7.5825t & -7.619t & -7.6594t & -7.6986t & {\bf -7.7329t} &\\\hline
	7 & -8.2239t & -8.2711t & -8.3296t & {\bf -8.351t} &&\\\hline
	6 & {\bf -8.5705t} & -8.502t & -8.4079t & -8.0969t &&\\\hline
	5 & {\bf -8.3087t} & -8.2013t & -7.7149t &&&\\\hline
	4 & -7.5317t & {\bf -7.5333t} & -7.0969t &&&\\\hline
	3 & {\bf -6.3387t} & -6.0969t &&&&\\\hline
	2 & {\bf -4.8299t} & -4.4788t &&&&\\\hline
\end{tabular}
\caption{Ground state energies of the 12-dot pentagon plaquette shown in Fig. \ref{fig:f5}.}
\end{table}

\newpage

\begin{figure}[H]
	\centering
	\includegraphics[width=.2\columnwidth]{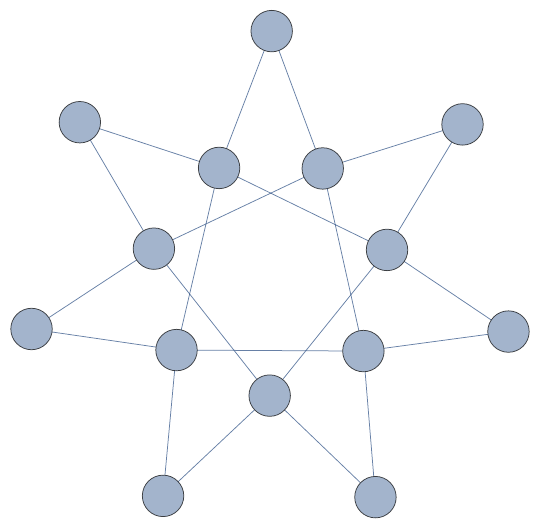}
	\includegraphics[width=.4\columnwidth]{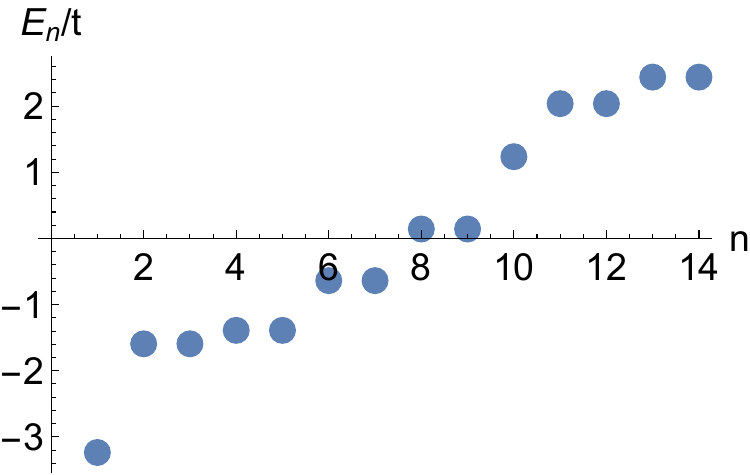}
		\caption{{\bf Left:} Graph of a 14-dot star-shaped plaquette comprised of pentagons. {\bf Right:} Single-particle spectrum of the 14-dot star-shaped pentagon plaquette.}
		\label{fig:f26}
\end{figure}

\begin{table}[H]
	\centering
\begin{tabular}{|c|c|c|c|c|c|c|c|}
	\hline
	\multicolumn{8}{|c|}{14 dots: Star-shaped pentagon graph}\\\hline
	& \multicolumn{7}{|c|}{Spin}\\\hline
	\# of el. & 0, 1/2 & 1, 3/2 & 2, 5/2 & 3, 7/2 & 4, 9/2 & 5, 11/2 & 6, 13/2\\\hline
	26 & -4.6516t & {\bf -4.8788t} &&&&&\\\hline
	25 & -6.843t & {\bf -6.9169t} &&&&&\\\hline
	24 & -8.3003t & -8.5319t & {\bf -8.955t} &&&&\\\hline
	23 & -9.6889t & -9.9269t & {\bf -10.1911t} &&&&\\\hline
	22 & {\bf -10.7263t} & -10.6943t & -10.6052t & -10.3337t &&&\\\hline
	21 & {\bf -10.8712t} & -10.8228t & -10.6194t & -10.4762t &&&\\\hline
	20 & {\bf -10.6737t} & -10.6383t & -10.5149t & -10.3202t & -9.8388t &&\\\hline
	19 & {\bf -10.1317t} & -10.0306t & -9.8912t & -9.5418t & -9.2013t &&\\\hline
	18 & {\bf -9.2059t} & -9.0941t & -8.9555t & -8.7185t & -8.4936t & -7.8118t &\\\hline
	17 & {\bf -7.5784t} & -7.5486t & -7.4442t & -7.3295t & -6.9062t & -6.4223t &\\\hline
	16 & -5.4834t & {\bf -5.5146t} & -5.4605t & -5.396t & -5.3068t & -5.1925t & -4.8292t \\\hline
	15 & -2.994t & -3.0349t & -3.0819t & -3.1251t & -3.1634t & -3.2004t & {\bf -3.2361t} \\\hline
	14 & 0 & 0 & 0 & 0 & 0 & 0 & 0 \\\hline
	13 & {\bf -3.0808t} & -3.0604t & -3.0238t & -2.9802t & -2.8432t & -2.7189t & -2.4394t \\\hline
	12 & {\bf -5.5364t} & -5.5334t & -5.5321t & -5.5309t & -5.4289t & -5.1191t & -4.8788t \\\hline
	11 & {\bf -7.5631t} & -7.5427t & -7.5198t & -7.4747t & -7.4352t & -6.9169t &\\\hline
	10 & -8.9555t & -8.9506t & -8.9781t & -9.0435t & {\bf -9.1168t} & -8.955t &\\\hline
	9 & -9.8684t & -9.9301t & -10.0162t & -10.1007t & {\bf -10.1911t} &&\\\hline
	8 & -10.2233t & -10.2606t & -10.357t & {\bf -10.4852t} & -10.3337t &&\\\hline
	7 & -10.2405t & -10.3396t & -10.411t & {\bf -10.4762t} &&&\\\hline
	6 & {\bf -10.1756t} & -10.0685t & -10.1432t & -9.8388t &&&\\\hline
	5 & {\bf -9.5293t} & -9.4899t & -9.2013t &&&&\\\hline
	4 & -8.4608t & {\bf -8.5781t} & -7.8118t &&&&\\\hline
	3 & {\bf -7.3018t} & -6.4223t &&&&&\\\hline
	2 & {\bf -6.002t} & -4.8292t &&&&&\\\hline
\end{tabular}
\caption{Ground state energies of the 14-dot star-shaped pentagon plaquette shown in Fig. \ref{fig:f26}.}
\end{table}

\newpage

\begin{figure}[H]
	\centering
	\includegraphics[width=.2\columnwidth]{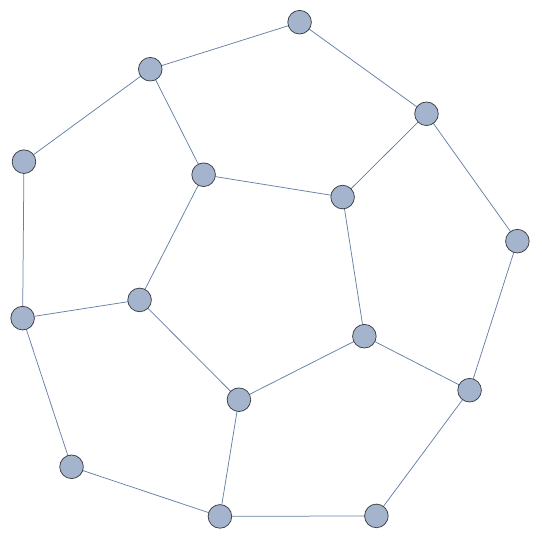}
	\includegraphics[width=.4\columnwidth]{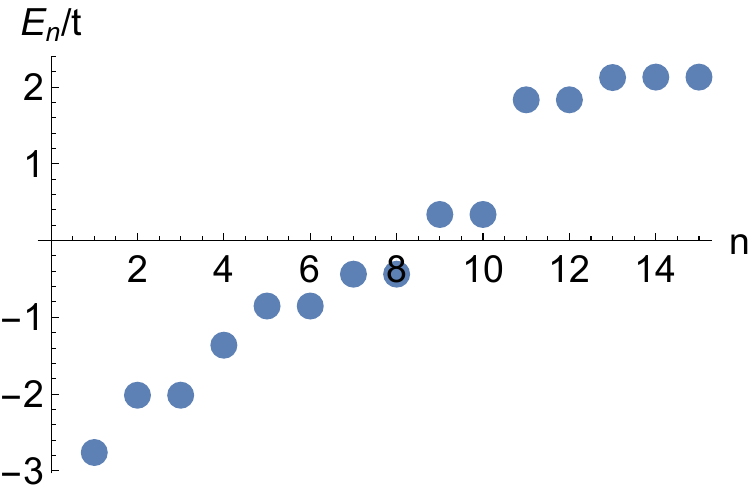}
	\caption{{\bf Left:} Graph of a 15-dot plaquette comprised of 6 adjacent pentagons. {\bf Right:} Single-particle spectrum of the 15-dot pentagon plaquette.}
	\label{fig:f17}
\end{figure}

\begin{table}[H]
	\centering
\begin{tabular}{|c|c|c|c|c|c|c|c|c|}
	\hline
	\multicolumn{9}{|c|}{15 dots: 6 adjacent pentagons}\\\hline
	& \multicolumn{8}{|c|}{Spin}\\\hline
	\# of el. & 0, 1/2 & 1, 3/2 & 2, 5/2 & 3, 7/2 & 4, 9/2 & 5, 11/2 & 6, 13/2 & 7\\\hline
	28 & -4.1656t & {\bf -4.261t} &&&&&&\\\hline
	27 & -6.2104t & {\bf -6.3859t} &&&&&&\\\hline
	26 & -8.041t & {\bf -8.241t} & -8.2206t &&&&&\\\hline
	25 & -9.7301t & -9.8536t & {\bf -10.0553t} &&&&&\\\hline
	24 & -10.5774t & -10.6633t & {\bf -10.7221t} & -10.3955t &&&&\\\hline
	23 & {\bf -11.0792t} & -11.0193t & -10.8981t & -10.7357t &&&&\\\hline
	22 & {\bf -11.1434t} & -11.0974t & -10.9806t & -10.8783t & -10.298t &&&\\\hline
	21 & {\bf -10.7828t} & -10.722t & -10.6327t & -10.1435t & -9.8604t &&&\\\hline
	20 & {\bf -10.0089t} & -9.921t & -9.8132t & -9.5486t & -9.3563t & -9.0077t &&\\\hline
	19 & {\bf -8.6727t} & -8.5827t & -8.4358t & -8.3604t & -8.2785t & -8.155t &&\\\hline
	18 & -6.9682t & {\bf -6.9991t} & -6.9221t & -6.8653t & -6.8325t & -6.8253t & -6.7917t &\\\hline
	17 & {\bf -4.9952t} & -4.9896t & -4.9808t & -4.9726t & -4.9694t & -4.9027t & -4.7766t &\\\hline
	16 & -2.6776t & -2.6881t & -2.701t & -2.7135t & -2.726t & -2.7386t & -2.7495t & {\bf -2.7616t} \\\hline
	15 & 0 & 0 & 0 & 0 & 0 & 0 & 0 & 0 \\\hline
	14 & {\bf -2.6851t} & -2.6818t & -2.6765t & -2.6681t & -2.6388t & -2.5725t & -2.4528t & -2.1305t \\\hline
	13 & -5.0292t & {\bf -5.0331t} & -5.0218t & -5.0206t & -4.9224t & -4.7281t & -4.261t &\\\hline
	12 & {\bf -6.9623t} & -6.9552t & -6.9534t & -6.956t & -6.9545t & -6.6837t & -6.3859t &\\\hline
	11 & -8.542t & -8.5681t & -8.5977t & -8.6152t & {\bf -8.6389t} & -8.2206t &&\\\hline
	10 & -9.8367t & -9.853t & -9.8985t & -9.9529t & -10.0067t & {\bf -10.0553t} &&\\\hline
	9 & -10.5t & -10.5615t & {\bf -10.5803t} & -10.5673t & -10.3955t &&&\\\hline
	8 & {\bf -10.9259t} & -10.8696t & -10.8431t & -10.8018t & -10.7357t &&&\\\hline
	7 & {\bf -10.9384t} & -10.793t & -10.7796t & -10.298t &&&&\\\hline
	6 & {\bf -10.6643t} & -10.4915t & -10.2417t & -9.8604t &&&&\\\hline
	5 & {\bf -9.7389t} & -9.633t & -9.0077t &&&&&\\\hline
	4 & -8.3722t & {\bf -8.5969t} & -8.155t &&&&&\\\hline
	3 & {\bf -6.9472t} & -6.7917t &&&&&&\\\hline
	2 & {\bf -5.2119t} & -4.7766t &&&&&&\\\hline
\end{tabular}
\caption{Ground state energies of the 15-dot pentagon plaquette shown in Fig. \ref{fig:f17}.}
\end{table}

\newpage

\begin{figure}[H]
	\centering
	\includegraphics[width=.2\columnwidth]{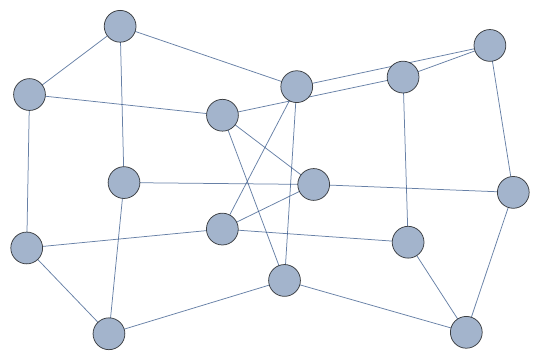}
	\includegraphics[width=.4\columnwidth]{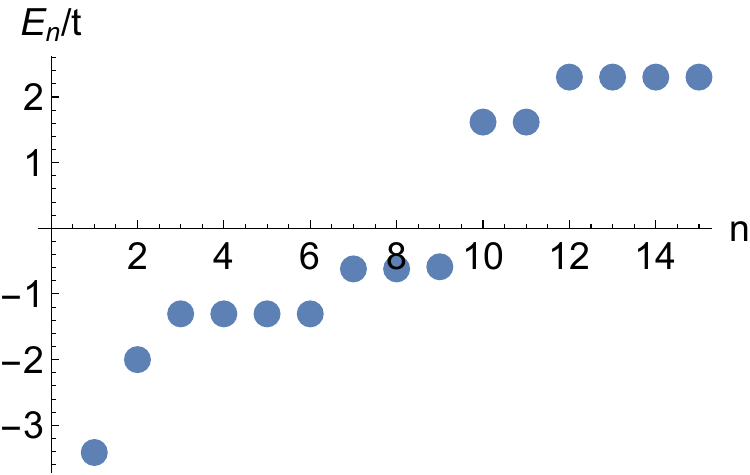}
	\caption{{\bf Left:} Graph of a 15-dot extended Petersen-graph-like plaquette. {\bf Right:} Single-particle spectrum of the 15-dot Petersen-like plaquette.}
	\label{fig:f28}
\end{figure}

\begin{table}[H]
	\centering
\begin{tabular}{|c|c|c|c|c|c|c|c|c|}
	\hline
	\multicolumn{9}{|c|}{15 dots: Extended Petersen-like graph}\\\hline
	& \multicolumn{8}{|c|}{Spin}\\\hline
	\# of el. & 0, 1/2 & 1, 3/2 & 2, 5/2 & 3, 7/2 & 4, 9/2 & 5, 11/2 & 6, 13/2 & 7\\\hline
	28 & -4.5578t & {\bf -4.6056t} &&&&&&\\\hline
	27 & -6.7961t & {\bf -6.9083t} &&&&&&\\\hline
	26 & -8.6315t & -8.8174t & {\bf -9.2111t} &&&&&\\\hline
	25 & -10.4787t & -10.6351t & {\bf -10.8291t} &&&&&\\\hline
	24 & -11.6656t & -12.0689t & -12.2841t & {\bf -12.4472t} &&&&\\\hline
	23 & -12.4858t & -12.5233t & {\bf -12.5833t} & -11.8614t &&&&\\\hline
	22 & {\bf -12.6648t} & -12.6154t & -12.5055t & -11.9606t & -11.2433t &&&\\\hline
	21 & {\bf -12.1192t} & -12.0669t & -11.6202t & -11.1259t & -10.6253t &&&\\\hline
	20 & {\bf -11.2825t} & -11.2616t & -10.9126t & -10.7145t & -10.1352t & -9.3225t &&\\\hline
	19 & {\bf -9.8989t} & -9.6739t & -9.4964t & -9.2168t & -8.8314t & -8.0198t &&\\\hline
	18 & {\bf -8.1578t} & -7.9353t & -7.8594t & -7.7094t & -7.4947t & -7.3014t & -6.717t &\\\hline
	17 & {\bf -5.7461t} & -5.7127t & -5.7355t & -5.6199t & -5.5861t & -5.5328t & -5.4142t &\\\hline
	16 & -3.128t & -3.1876t & -3.2409t & -3.2798t & -3.3158t & -3.35t & -3.3827t & {\bf -3.4142t} \\\hline
	15 & 0 & 0 & 0 & 0 & 0 & 0 & 0 & 0 \\\hline
	14 & {\bf -3.1551t} & -3.1446t & -3.1343t & -3.0957t & -3.0657t & -2.9451t & -2.7321t & -2.3028t \\\hline
	13 & {\bf -5.7321t} & -5.7071t & -5.6682t & -5.5907t & -5.5169t & -5.1702t & -4.6056t &\\\hline
	12 & {\bf -7.9853t} & -7.9708t & -7.9409t & -7.899t & -7.8151t & -7.6481t & -6.9083t &\\\hline
	11 & -9.6699t & -9.7094t & -9.7719t & {\bf -9.8425t} & -9.5661t & -9.2111t &&\\\hline
	10 & -11.0719t & -11.0686t & -11.1806t & -11.2834t & {\bf -11.3775t} & -10.8291t &&\\\hline
	9 & -11.7586t & -12.0485t & -12.1864t & -12.321t & {\bf -12.4472t} &&&\\\hline
	8 & -11.7941t & -11.8133t & -11.957t & {\bf -12.0985t} & -11.8614t &&&\\\hline
	7 & -11.7435t & {\bf -11.8323t} & -11.7768t & -11.2433t &&&&\\\hline
	6 & -11.1132t & {\bf -11.2399t} & -11.0298t & -10.6253t &&&&\\\hline
	5 & {\bf -10.3743t} & -10.2175t & -9.3225t &&&&&\\\hline
	4 & {\bf -9.533t} & -9.1764t & -8.0198t &&&&&\\\hline
	3 & {\bf -8.1235t} & -6.717t &&&&&&\\\hline
	2 & {\bf -6.3917t} & -5.4142t &&&&&&\\\hline
\end{tabular}
\caption{Ground state energies of the 15-dot extended petersen-like plaquette shown in Fig. \ref{fig:f28}.}
\end{table}

\newpage

\begin{figure}[H]
	\centering
	\includegraphics[width=.2\columnwidth]{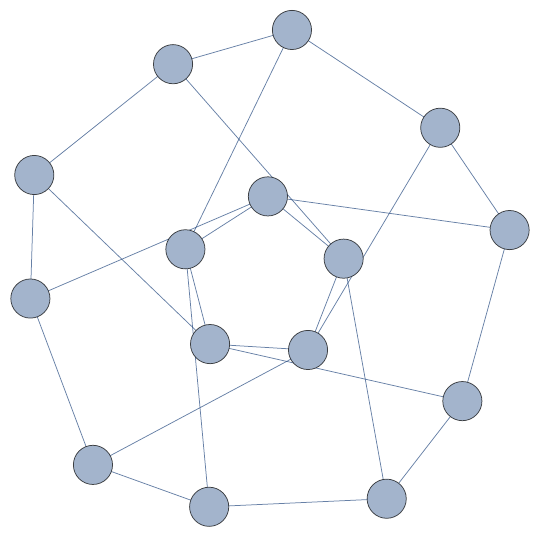}
	\includegraphics[width=.4\columnwidth]{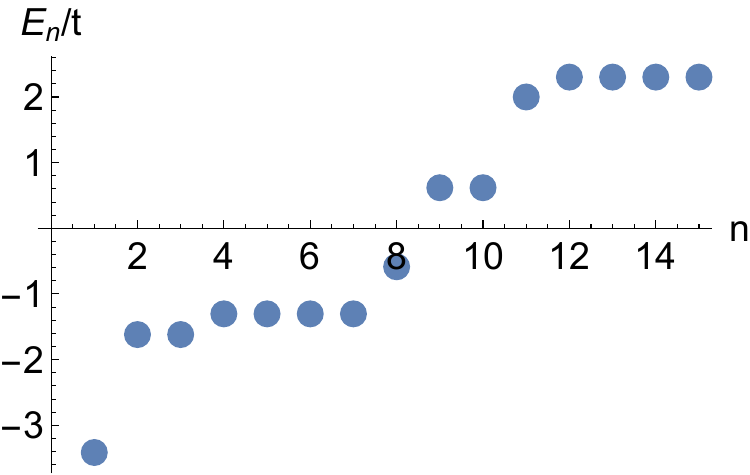}
	\caption{{\bf Left:} Graph of a 15-dot interconnected pentagon plaquette. {\bf Right:} Single-particle spectrum of the 15-dot interconnected pentagon plaquette.}
	\label{fig:f27}
\end{figure}

\begin{table}[H]
	\centering
\begin{tabular}{|c|c|c|c|c|c|c|c|c|}
	\hline
	\multicolumn{9}{|c|}{15 dots: interconnected pentagon plaquette}\\\hline
	& \multicolumn{8}{|c|}{Spin}\\\hline
	\# of el. & 0, 1/2 & 1, 3/2 & 2, 5/2 & 3, 7/2 & 4, 9/2 & 5, 11/2 & 6, 13/2 & 7\\\hline
	28 & -4.5578t & {\bf -4.6056t} &&&&&&\\\hline
	27 & -6.7942t & {\bf -6.9083t} &&&&&&\\\hline
	26 & -8.6338t & -8.8308t & {\bf -9.2111t} &&&&&\\\hline
	25 & -10.5143t & -10.7091t & {\bf -11.2111t} &&&&&\\\hline
	24 & -11.5908t & -11.9746t & {\bf -12.0824t} & -11.8291t &&&&\\\hline
	23 & -12.4735t & {\bf -12.5575t} & -12.4857t & -12.4472t &&&&\\\hline
	22 & -12.5256t & {\bf -12.5462t} & -12.4654t & -12.3996t & -11.8614t &&&\\\hline
	21 & {\bf -12.1594t} & -12.0903t & -12.1075t & -11.4736t & -10.5586t &&&\\\hline
	20 & {\bf -11.2943t} & -11.1636t & -11.1431t & -10.7053t & -10.1749t & -9.2558t &&\\\hline
	19 & {\bf -9.8447t} & -9.8206t & -9.4717t & -9.1139t & -8.6654t & -7.9531t &&\\\hline
	18 & -8.0126t & {\bf -8.0651t} & -7.8529t & -7.6361t & -7.4074t & -7.0808t & -6.6503t &\\\hline
	17 & {\bf -5.7649t} & -5.7158t & -5.6139t & -5.5336t & -5.4402t & -5.3162t & -5.0322t &\\\hline
	16 & -3.0676t & -3.122t & -3.187t & -3.2314t & -3.2796t & -3.3269t & -3.372t & {\bf -3.4142t} \\\hline
	15 & 0 & 0 & 0 & 0 & 0 & 0 & 0 & 0 \\\hline
	14 & {\bf -3.1482t} & -3.1447t & -3.1318t & -3.1036t & -3.0661t & -2.9458t & -2.7321t & -2.3028t \\\hline
	13 & {\bf -5.7276t} & -5.7142t & -5.6867t & -5.6525t & -5.5176t & -5.171t & -4.6056t &\\\hline
	12 & {\bf -7.9645t} & -7.955t & -7.9477t & -7.9326t & -7.8926t & -7.6479t & -6.9083t &\\\hline
	11 & -9.6185t & -9.6901t & -9.7601t & -9.7642t & {\bf -9.824t} & -9.2111t &&\\\hline
	10 & -10.9797t & -10.9955t & -11.1103t & -11.1181t & -11.1385t & {\bf -11.2111t} &&\\\hline
	9 & -11.7087t & -11.8905t & -12.0485t & {\bf -12.1421t} & -11.8291t &&&\\\hline
	8 & -11.9041t & -11.9978t & -12.1455t & -12.3799t & {\bf -12.4472t} &&&\\\hline
	7 & -11.5821t & -11.7688t & {\bf -11.9279t} & -11.8614t &&&&\\\hline
	6 & -10.9339t & -10.9717t & {\bf -11.187t} & -10.5586t &&&&\\\hline
	5 & -10.0513t & {\bf -10.1728t} & -9.2558t &&&&&\\\hline
	4 & -8.8997t & {\bf -9.1522t} & -7.9531t &&&&&\\\hline
	3 & {\bf -7.7868t} & -6.6503t &&&&&&\\\hline
	2 & {\bf -6.39t} & -5.0322t &&&&&&\\\hline
\end{tabular}
\caption{Ground state energies of the 15-dot interconnected pentagon plaquette shown in Fig. \ref{fig:f27}.}
\end{table}

\newpage

\subsection{Other Plaquette Geometries}

\begin{figure}[H]
	\centering
	\includegraphics[width=.2\columnwidth]{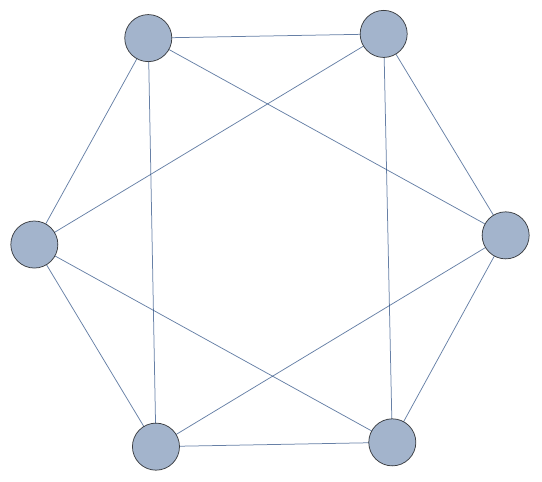}
	\includegraphics[width=.4\columnwidth]{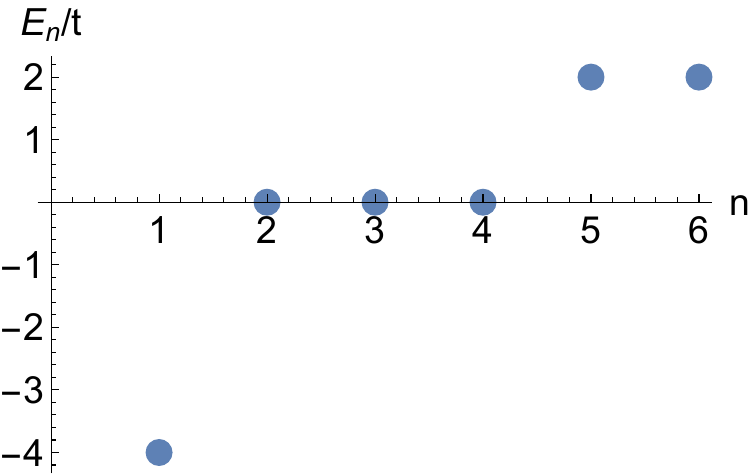}
	\caption{{\bf Left:} Graph of a 6-dot octahedron plaquette. {\bf Right:} Single-particle spectrum of the  6-dot octahedron plaquette.}
	\label{fig:fs8}
\end{figure}

\begin{table}[H]
	\centering
\begin{tabular}{|c|c|c|c|c|}
	\hline
	\multicolumn{5}{|c|}{6 dots: octahedron}\\\hline
	& \multicolumn{4}{|c|}{Spin}\\\hline
	\# of el. & 0, 1/2 & 1, 3/2 & 2, 5/2 & 3 \\\hline
	10 & -3.236t -0.724J  & {\bf -4t } &&\\\hline
	9 & {\bf -4.509t -1.088J }  & -4t &&\\\hline
	8 & {\bf -5.123t -1.894J }  & -4.503t -1.102J  & -4t &\\\hline
	7 & -3t -2.55J  & -3.449t -1.638J  & {\bf -4t } &\\\hline
	6 & {\bf -6J }  & -5J  & -3J  & 0 \\\hline
	5 & {\bf -3t -5.75J }  & -3t -4.25J  & -2t &\\\hline
	4 & -4.275t -4.947J  & {\bf -4.302t -3.867J }  & -4t &\\\hline
	3 & {\bf -5.413t -3.573J }  & -4t &&\\\hline
	2 & {\bf -6.472t -2.894J }  & -4t &&\\\hline
\end{tabular}
	\caption{Ground state energies of the 6-dot octahedron plaquette shown in Fig. \ref{fig:fs8}.}
\end{table}

\newpage

\begin{figure}[H]
	\centering
	\includegraphics[width=.2\columnwidth]{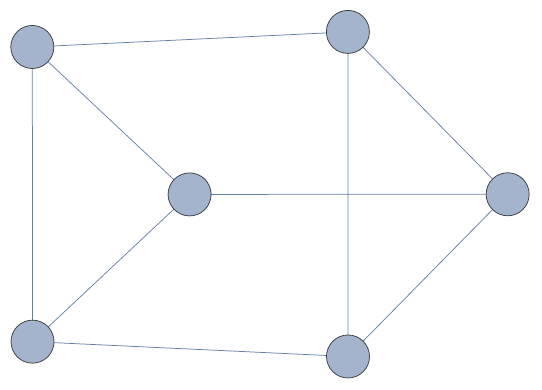}
	\includegraphics[width=.4\columnwidth]{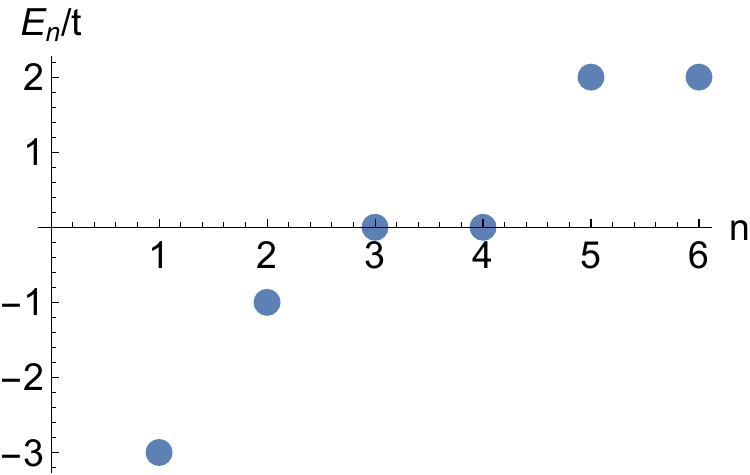}
	\caption{{\bf Left:} Graph of a 6-dot triangular prism plaquette. {\bf Right:} Single-particle spectrum of the 6-dot triangular prism plaquette.}
	\label{fig:fs9}
\end{figure}

\begin{table}[H]
	\centering
	\begin{tabular}{|c|c|c|c|c|}
		\hline
		\multicolumn{5}{|c|}{6 dots: triangular prism}\\\hline
		& \multicolumn{4}{|c|}{Spin}\\\hline
		\# of el. & 0, 1/2 & 1, 3/2 & 2, 5/2 & 3 \\\hline
		10 & -3.323t -0.556J  & {\bf -4t } &&\\\hline
		9 & {\bf -4.39t -1.305J }  & -4t &&\\\hline
		8 & {\bf -4.614t -2.138J }  & -4.459t -1.207J  & -4t &\\\hline
		7 & -2.562t -1.852J  & -2.791t -1.173J  & {\bf -3t } &\\\hline
		6 & {\bf -5.303J }  & -4.281J  & -2.5J  & 0 \\\hline
		5 & {\bf -2.732t -3.684J }  & -2.618t -3.171J  & -2t &\\\hline
		4 & {\bf -4.152t -4.126J }  & -4.07t -2.284J  & -4t &\\\hline
		3 & {\bf -4.766t -2.509J }  & -4t &&\\\hline
		2 & {\bf -4.962t -1.306J }  & -4t &&\\\hline
	\end{tabular}
	\caption{Ground state energies of the 6-dot triangular prism plaquette shown in Fig. \ref{fig:fs9}.}
\end{table}

\newpage

\begin{figure}[H]
	\centering
	\includegraphics[width=.2\columnwidth]{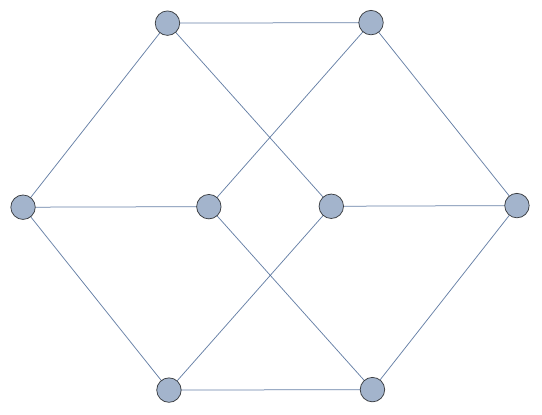}
	\includegraphics[width=.4\columnwidth]{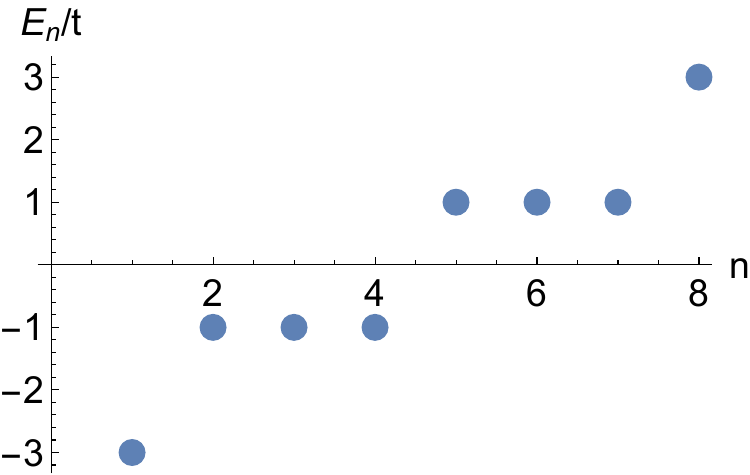}
	\caption{{\bf Left:} Graph of a 8-dot cubic plaquette. {\bf Right:} Single-particle spectrum of the 8-dot cubic plaquette.}
	\label{fig:fs10}
\end{figure}

\begin{table}[H]
	\centering
	\begin{tabular}{|c|c|c|c|c|c|}
		\hline
		\multicolumn{6}{|c|}{8 dots: cube}\\\hline
		& \multicolumn{5}{|c|}{Spin}\\\hline
		\# of el. & 0, 1/2 & 1, 3/2 & 2, 5/2 & 3, 7/2 & 4 \\\hline
		8 & {\bf -7.82J }  & -7J  & -5.414J  & -3J  & 0 \\\hline
		7, 9 & -2.625t -3.573J  & -2.714t -1.636J  & -2.858t -1.124J  & {\bf -3t } &\\\hline
		6, 10 & {\bf -4.788t -1.811J }  & -4.673t -2.032J  & -4.51t -1.092J & -4t &\\\hline
		5, 11 & -5.546t -3.056J  & {\bf -5.984t -2.47J }  & -5t &&\\\hline
		4, 12 & -5.962t -2.157J  & -5.951t -1.717J  & {\bf -6t } &&\\\hline
		3, 13 & {\bf -5.658t -1.329J }  & -5t &&&\\\hline
		2, 14 & {\bf -5.292t -0.857J }  & -4t &&&\\\hline
	\end{tabular}
	\caption{Ground state energies of the 8-dot cubic plaquette shown in Fig. \ref{fig:fs10}.}
\end{table}

\newpage

\begin{figure}[H]
	\centering
	\includegraphics[width=.2\columnwidth]{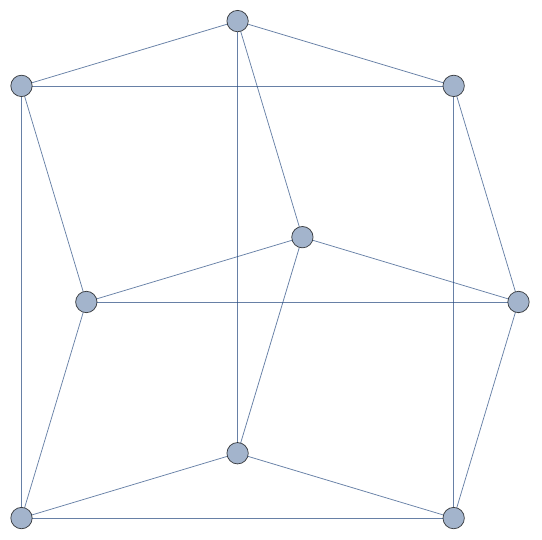}
	\includegraphics[width=.4\columnwidth]{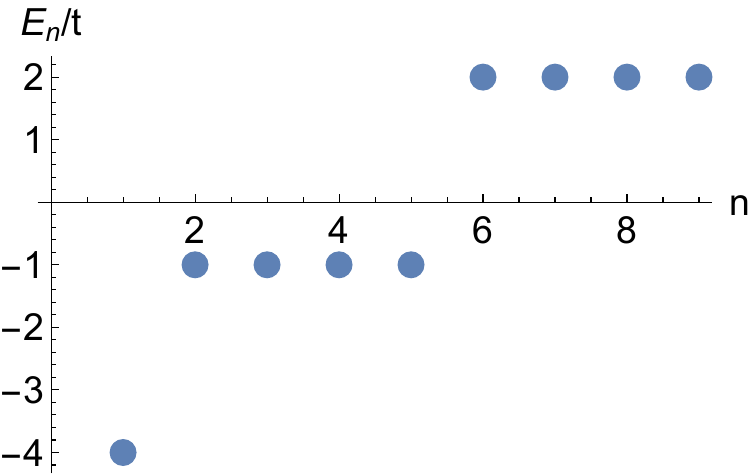}
	\caption{{\bf Left:} Graph of a 9-dot plaquette arrange in a (2,3)-Hamming graph. {\bf Right:} Single-particle spectrum of the 9-dot (2,3)-Hamming graph.}
	\label{fig:f0}
\end{figure}

\begin{table}[H]
	\centering
	\begin{tabular}{|c|c|c|c|c|c|}
		\hline
		\multicolumn{6}{|c|}{9 dots: (2,3)-Hamming graph}\\\hline
		& \multicolumn{5}{|c|}{Spin}\\\hline
		\# of el. & 0, 1/2 & 1, 3/2 & 2, 5/2 & 3, 7/2 & 4\\\hline
		16 & {\bf -4t} & {\bf -4t} &&&\\\hline
		15 & -5.5857t & {\bf -6t} &&&\\\hline
		14 & -7.1945t & -7.1315t & {\bf -8t} &&\\\hline
		13 & -7.4647t & {\bf -7.7421t} & -7t &&\\\hline
		12 & {\bf -7.7014t} & -7.2919t & -6.8076t & -6t &\\\hline
		11 & {\bf -5.9511t} & -5.9366t & -5.8503t & -5t &\\\hline
		10 & -3.1623t & -3.3456t & -3.5616t & -3.7762t & {\bf -4t} \\\hline
		9 & 0 & 0 & 0 & 0 & 0 \\\hline
		8 & {\bf -3.4142t} & -3.3162t & -3.1009t & -2.7321t & -2t \\\hline
		7 & -5.5185t & {\bf -5.534t} & -5.063t & -4t &\\\hline
		6 & -6.7831t & -6.9136t & {\bf -7.4357t} & -6t &\\\hline
		5 & -7.507t & -7.9138t & {\bf -8t} &&\\\hline
		4 & -7.5146t & {\bf -7.8541t} & -7t &&\\\hline
		3 & {\bf -7.4774t} & -6t &&&\\\hline
		2 & {\bf -7.1231t} & -5t &&&\\\hline
	\end{tabular}
	\caption{Ground state energies of the 9-dot (2,3)-Hamming graph shown in Fig. \ref{fig:f0}.}
\end{table}

\newpage

\begin{figure}[H]
	\centering
	\includegraphics[width=.2\columnwidth]{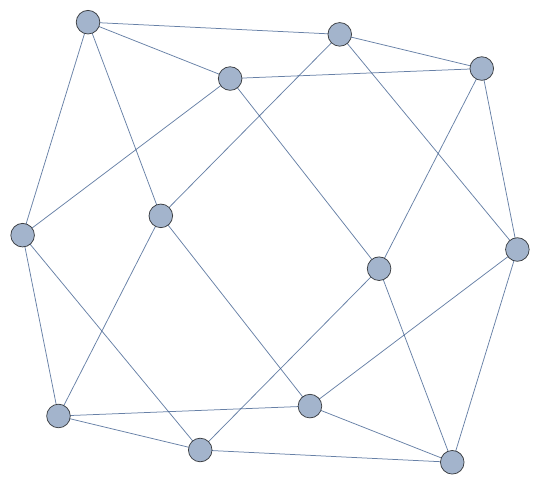}
	\includegraphics[width=.4\columnwidth]{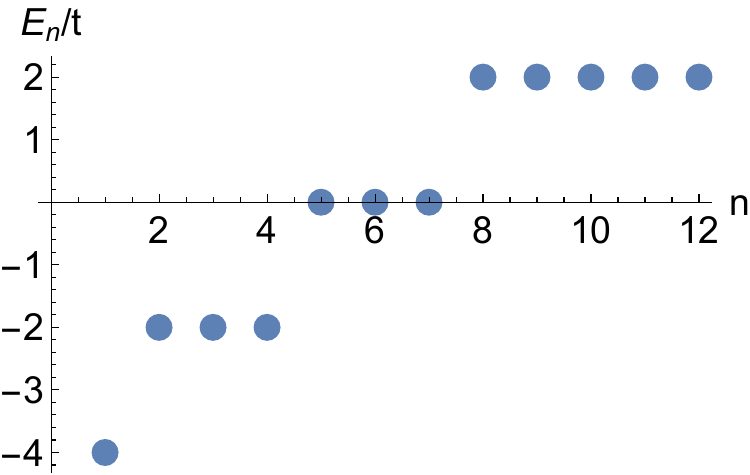}
	\caption{{\bf Left:} Graph of a 12-dot cuboctahedron plaquette. {\bf Right:} Single-particle spectrum of the 12-dot cuboctahedron plaquette.}
	\label{fig:f7}
\end{figure}

\begin{table}[H]
	\centering
	\begin{tabular}{|c|c|c|c|c|c|c|}
		\hline
		\multicolumn{7}{|c|}{12 dots: cuboctahedron}\\\hline
		& \multicolumn{6}{|c|}{Spin}\\\hline
		\# of el. & 0, 1/2 & 1, 3/2 & 2, 5/2 & 3, 7/2 & 4, 9/2 & 5, 11/2\\\hline
		22 & {\bf -4t} & {\bf -4t} &&&&\\\hline
		21 & -5.9153t & {\bf -6t} &&&&\\\hline
		20 & -7.7375t & -7.7468t & {\bf -8t} &&&\\\hline
		19 & -9.2002t & -9.451t & {\bf -10t} &&&\\\hline
		18 & {\bf -10.3255t} & -10.1679t & -10.2674t & -10t &&\\\hline
		17 & -10.2328t & -10.3376t & {\bf -10.3892t} & -10t &&\\\hline
		16 & -9.9934t & -9.9072t & -9.7962t & -9.7505t & {\bf -10t} &\\\hline
		15 & -8.3886t & {\bf -8.471t} & -8.4547t & -8.4338t & -8t &\\\hline
		14 & {\bf -6.6394t} & -6.5918t & -6.534t & -6.5102t & -6.4667t & -6t \\\hline
		13 & -3.5224t & -3.6096t & -3.7088t & -3.809t & -3.9058t & {\bf -4t} \\\hline
		12 & 0 & 0 & 0 & 0 & 0 & 0 \\\hline
		11 & {\bf -3.4943t} & -3.4683t & -3.2978t & -3.0755t & -2.7913t & -2t \\\hline
		10 & {\bf -6.0941t} & -6.0754t & -6.0063t & -5.5833t & -5.0187t & -4t \\\hline
		9 & {\bf -8.2176t} & -8.1762t & -7.9333t & -7.1942t & -6t &\\\hline
		8 & {\bf -10.1759t} & -9.6179t & -9.3573t & -9.3707t & -8t &\\\hline
		7 & {\bf -10.9403t} & -10.5902t & -10.2t & -10t &&\\\hline
		6 & -11.2782t & {\bf -11.3335t} & -10.8622t & -10t &&\\\hline
		5 & -10.9465t & {\bf -11.4593t} & -10t &&&\\\hline
		4 & -10.0423t & {\bf -10.2462t} & -10t &&&\\\hline
		3 & {\bf -8.8681t} & -8t &&&&\\\hline
		2 & {\bf -7.4186t} & -6t &&&&\\\hline
	\end{tabular}
	\caption{Ground state energies of the 12-dot cuboctahedron plaquette shown in Fig. \ref{fig:f7}.}
\end{table}

\newpage

\begin{figure}[H]
	\centering
	\includegraphics[width=.2\columnwidth]{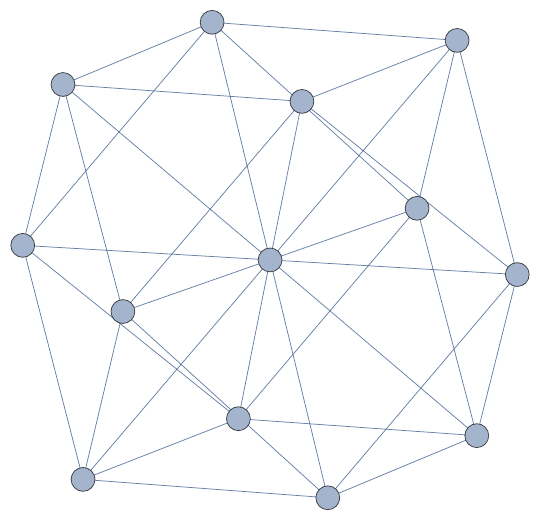}
	\includegraphics[width=.4\columnwidth]{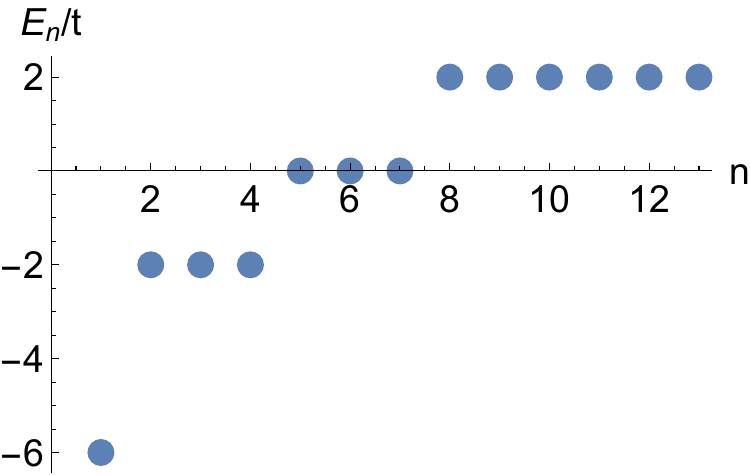}
	\caption{{\bf Left:} Graph of a 13-dot subsection of an FCC-lattice. {\bf Right:} Single-particle spectrum of the 13-dot FCC-lattice.}
	\label{fig:f10}
\end{figure}

\begin{table}[H]
	\centering
	\begin{tabular}{|c|c|c|c|c|c|c|c|}
		\hline
		\multicolumn{8}{|c|}{13 dots: FCC lattice}\\\hline
		& \multicolumn{7}{|c|}{Spin}\\\hline
		\# of el. & 0, 1/2 & 1, 3/2 & 2, 5/2 & 3, 7/2 & 4, 9/2 & 5, 11/2 & 6\\\hline
		24 & {\bf -4t} & {\bf -4t} &&&&&\\\hline
		23 & {\bf -6t} & {\bf -6t} &&&&&\\\hline
		22 & -7.7403t & -7.871t & {\bf -8t} &&&&\\\hline
		21 & -9.5552t & -9.6625t & {\bf -10t} &&&&\\\hline
		20 & -10.8105t & -11.0213t & -11.3645t & {\bf -12t} &&&\\\hline
		19 & -12.0246t & -11.988t & {\bf -12.1859t} & -12t &&&\\\hline
		18 & -11.7258t & -11.9205t & -12.1634t & {\bf -12.2829t} & -12t &&\\\hline
		17 & -11.5713t & -11.5946t & -11.6016t & -11.6503t & {\bf -12t} &&\\\hline
		16 & -9.7317t & -9.9794t & -10.1818t & -10.2407t & {\bf -10.3303t} & -10t &\\\hline
		15 & -8.0492t & -8.135t & -8.2188t & -8.3152t & {\bf -8.3707t} & -8t &\\\hline
		14 & -5.3755t & -5.3902t & -5.4543t & -5.5604t & -5.6951t & -5.8436t & {\bf -6t} \\\hline
		13 & 0 & 0 & 0 & 0 & 0 & 0 & 0 \\\hline
		12 & {\bf -3.7243t} & -3.7077t & -3.646t & -3.5509t & -3.3313t & -2.9438t & -2t \\\hline
		11 & {\bf -7.1654t} & -7.1024t & -6.7447t & -6.2969t & -5.6731t & -4t &\\\hline
		10 & {\bf -9.6813t} & -9.637t & -9.535t & -8.9292t & -8.1277t & -6t &\\\hline
		9 & {\bf -11.7707t} & -11.701t & -11.395t & -10.4625t & -8t &&\\\hline
		8 & {\bf -13.7166t} & -13.0744t & -12.7837t & -12.8213t & -10t &&\\\hline
		7 & {\bf -14.3472t} & -13.9839t & -13.5181t & -12t &&&\\\hline
		6 & -14.546t & {\bf -14.6431t} & -14.1514t & -12t &&&\\\hline
		5 & -14.2217t & {\bf -14.748t} & -12t &&&&\\\hline
		4 & -13.2602t & {\bf -13.4679t} & -12t &&&&\\\hline
		3 & {\bf -12.0947t} & -10t &&&&&\\\hline
		2 & {\bf -10.6671t} & -8t &&&&&\\\hline
	\end{tabular}
	\caption{Ground state energies of the 13-dot FCC-lattice shown in Fig. \ref{fig:f10}.}
\end{table}

\newpage

\begin{figure}[H]
	\centering
	\includegraphics[width=.2\columnwidth]{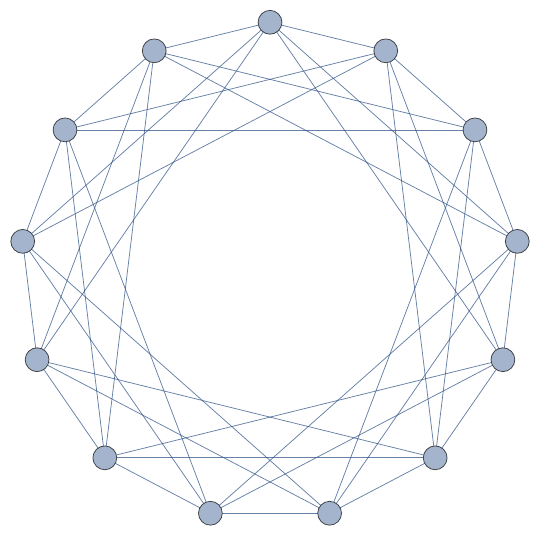}
	\includegraphics[width=.2\columnwidth]{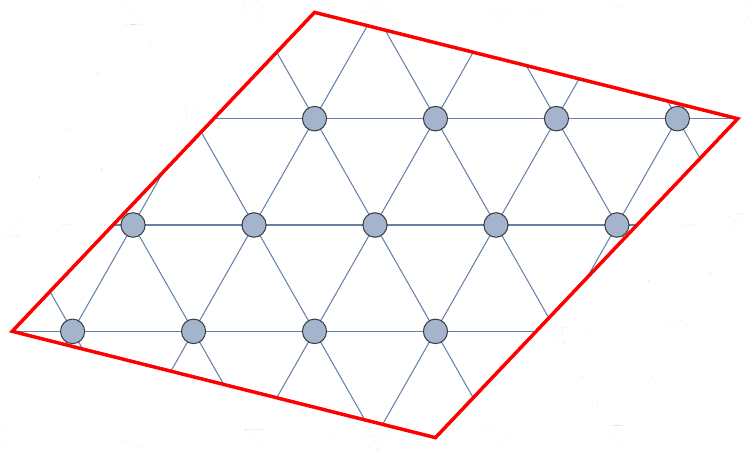}
	\includegraphics[width=.4\columnwidth]{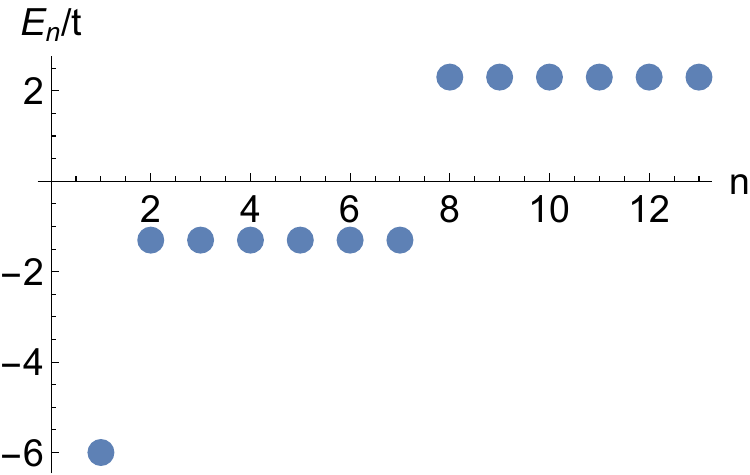}
	\caption{{\bf Left:} Graph of the Paley-13 graph. {\bf Middle:} Paley-13 graph represented as a triangle lattice with periodic boundary conditions. {\bf Right:} Single-particle spectrum of the  Paley-13 graph plaquette.}
	\label{fig:f13}
\end{figure}

\begin{table}[H]
	\centering
	\begin{tabular}{|c|c|c|c|c|c|c|c|}
		\hline
		\multicolumn{8}{|c|}{13 dots: Paley-13 graph}\\\hline
		& \multicolumn{7}{|c|}{Spin}\\\hline
		\# of el. & 0, 1/2 & 1, 3/2 & 2, 5/2 & 3, 7/2 & 4, 9/2 & 5, 11/2 & 6\\\hline
		24 & {\bf -4.6056t} & {\bf -4.6056t} &&&&&\\\hline
		23 & {\bf -6.9083t} & {\bf -6.9083t} &&&&&\\\hline
		22 & -8.7177t & -8.9812t & {\bf -9.2111t} &&&&\\\hline
		21 & -10.7043t & -10.9538t & {\bf -11.5139t} &&&&\\\hline
		20 & -11.79t & -12.1959t & -12.7335t & {\bf -13.8167t} &&&\\\hline
		19 & -12.8266t & -12.9854t & {\bf -13.4368t} & -12.5139t &&&\\\hline
		18 & -12.6951t & -12.8908t & {\bf -13.0302t} & -12.1777t & -11.2111t &&\\\hline
		17 & {\bf -12.3798t} & -12.1954t & -11.618t & -10.8536t & -9.9083t &&\\\hline
		16 & -10.784t & {\bf -10.7897t} & -10.5095t & -10.1266t & -9.4571t & -8.6056t &\\\hline
		15 & {\bf -8.3931t} & -8.3615t & -8.3062t & -8.2131t & -7.9285t & -7.3028t &\\\hline
		14 & -4.4142t & -4.629t & -4.9018t & -5.1824t & -5.4527t & -5.7152t & {\bf -6t} \\\hline
		13 & 0 & 0 & 0 & 0 & 0 & 0 & 0 \\\hline
		12 & {\bf -4.1205t} & -4.1019t & -3.9503t & -3.8287t & -3.5009t & -3.2361t & -2.3028t \\\hline
		11 & {\bf -7.5209t} & -7.4412t & -7.0293t & -6.5704t & -5.8856t & -4.6056t &\\\hline
		10 & -9.8407t & {\bf -9.9917t} & -9.9107t & -9.2735t & -8.6215t & -6.9083t &\\\hline
		9 & -11.498t & -11.7485t & {\bf -12.0227t} & -11.2606t & -9.2111t &&\\\hline
		8 & -12.6485t & -12.819t & -13.2218t & {\bf -13.9952t} & -11.5139t &&\\\hline
		7 & -13.2178t & -13.6182t & {\bf -14.034t} & -13.8167t &&&\\\hline
		6 & -13.3629t & -13.3112t & {\bf -13.8504t} & -12.5139t &&&\\\hline
		5 & -13.0425t & {\bf -13.233t} & -11.2111t &&&&\\\hline
		4 & -12.5132t & {\bf -12.5738t} & -9.9083t &&&&\\\hline
		3 & {\bf -11.773t} & -8.6056t &&&&&\\\hline
		2 & {\bf -11.0828t} & -7.3028t &&&&&\\\hline
	\end{tabular}
	\caption{Ground state energies of the Paley-13 graph plaquette shown in Fig. \ref{fig:f13}.}
\end{table}

\newpage

\begin{figure}[H]
	\centering
	\includegraphics[width=.2\columnwidth]{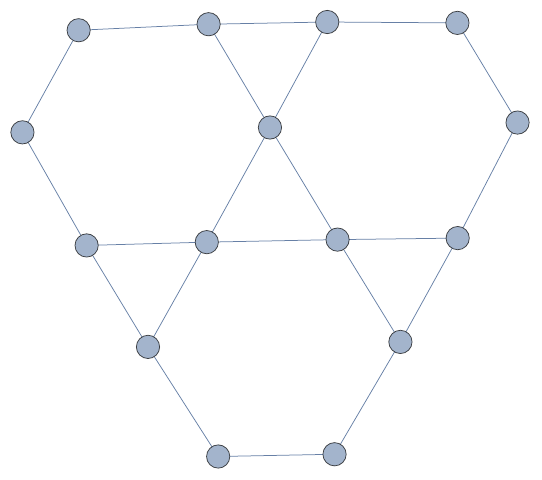}
	\includegraphics[width=.4\columnwidth]{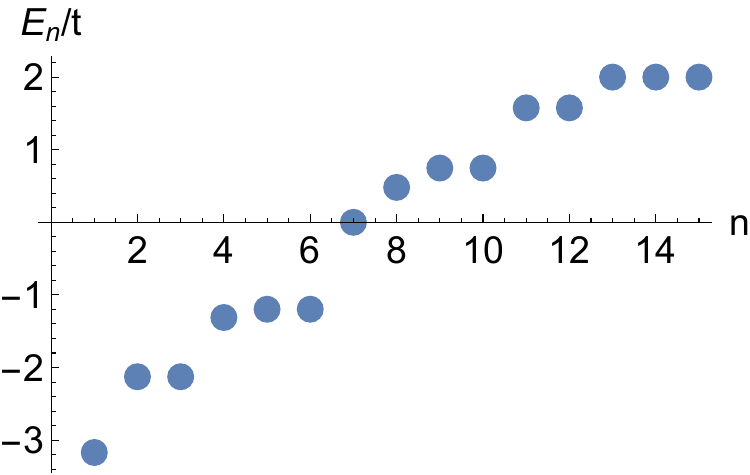}
	\caption{{\bf Left:} Graph of a 15-dot Kogome lattice. {\bf Right:} Single-particle spectrum of the 15-dot Kogome lattice.}
	\label{fig:f18}
\end{figure}

\begin{table}[H]
	\centering
	\begin{tabular}{|c|c|c|c|c|c|c|c|c|}
		\hline
		\multicolumn{9}{|c|}{15 dots: Kagome lattice}\\\hline
		& \multicolumn{8}{|c|}{Spin}\\\hline
		\# of el. & 0, 1/2 & 1, 3/2 & 2, 5/2 & 3, 7/2 & 4, 9/2 & 5, 11/2 & 6, 13/2 & 7\\\hline
		28 & -3.9277t & {\bf -4t} &&&&&&\\\hline
		27 & -5.8787t & {\bf -6t} &&&&&&\\\hline
		26 & -7.4213t & {\bf -7.5991t} & -7.5764t &&&&&\\\hline
		25 & -8.9092t & -9.0222t & {\bf -9.1528t} &&&&&\\\hline
		24 & {\bf -10.1696t} & -10.11t & -10.1066t & -9.9003t &&&&\\\hline
		23 & -10.7478t & {\bf -10.8128t} & -10.7663t & -10.6478t &&&&\\\hline
		22 & -10.9083t & -11.0028t & -11.0315t & -11.0732t & {\bf -11.129t} &&&\\\hline
		21 & -10.8369t & -10.9169t & -10.9944t & -11.0684t & {\bf -11.129t} &&&\\\hline
		20 & -10.0649t & -10.0992t & -10.1195t & {\bf -10.1389t} & -10.1282t & -9.9318t &&\\\hline
		19 & -8.9526t & -8.9717t & {\bf -8.9863t} & -8.9146t & -8.8603t & -8.7347t &&\\\hline
		18 & -7.5772t & -7.5839t & -7.5922t & {\bf -7.5996t} & -7.5299t & -7.4605t & -7.4236t &\\\hline
		17 & -5.4241t & -5.4286t & -5.4323t & {\bf -5.4351t} & -5.4267t & -5.4108t & -5.2968t &\\\hline
		16 & -3.1477t & -3.1506t & -3.1532t & -3.1568t & -3.159t & -3.1628t & -3.1657t & {\bf -3.1701t} \\\hline
		15 & 0 & 0 & 0 & 0 & 0 & 0 & 0 & 0 \\\hline
		14 & {\bf -3.1363t} & {\bf -3.1362t} & -3.1361t & -3.136t & -3.0327t & -2.8955t & -2.6492t & -2t \\\hline
		13 & -5.4285t & {\bf -5.4336t} & -5.4327t & -5.4302t & -5.2029t & -4.8018t & -4t &\\\hline
		12 & {\bf -7.5739t} & -7.5375t & -7.5017t & -7.4698t & -7.0525t & -6.6443t & -6t &\\\hline
		11 & {\bf -9.0721t} & -9.0462t & -8.9431t & -8.6048t & -8.4393t & -7.5764t &&\\\hline
		10 & -10.3647t & -10.3502t & {\bf -10.3714t} & -9.8731t & -9.514t & -9.1528t &&\\\hline
		9 & {\bf -11.3738t} & -11.3691t & -11.0776t & -10.5407t & -9.9003t &&&\\\hline
		8 & {\bf -11.7808t} & -11.752t & -11.7035t & -11.4988t & -10.6478t &&&\\\hline
		7 & -11.6469t & -11.6703t & {\bf -11.6974t} & -11.129t &&&&\\\hline
		6 & {\bf -11.2495t} & -11.1487t & -11.1253t & -11.129t &&&&\\\hline
		5 & {\bf -10.3727t} & -10.2133t & -9.9318t &&&&&\\\hline
		4 & -9.023t & {\bf -9.2046t} & -8.7347t &&&&&\\\hline
		3 & {\bf -7.6089t} & -7.4236t &&&&&&\\\hline
		2 & {\bf -5.74t} & -5.2968t &&&&&&\\\hline
	\end{tabular}
	\caption{Ground state energies of the 15-dot Kogome lattice shown in Fig. \ref{fig:f18}.}
\end{table}

\end{widetext}

\bibliography{magnetbib}

\end{document}